\documentclass[12pt]{iopart}
\pdfoutput=1

\usepackage{graphicx}
\usepackage{color}
\usepackage{epsfig}

\usepackage{bm}        
\usepackage{amsbsy}
\usepackage{upgreek}

\newcommand{\nvec}[1]{{\rm \bf #1}}

\begin{document}
\title[]{Visualization and interpretation of Rydberg states}

\author{Ladislav Kocbach$\dagger$ and Abdul Waheed$\dagger$\P}


\address{$\dagger$Department of Physics and Technology, University of Bergen, N-5007 Bergen, Norway}
\address{\P\ Higher Education Commission of Pakistan, Islamabad, Pakistan}

\ead{ladislav.kocbach@ift.uib.no, \ waheed.tanoli@gmail.com }

\begin{abstract}
For many purposes it is desirable to have an easily understandable and accurate 
picture of the atomic states. This is especially true for the highly excited states which exhibit features 
not present in the well known states hydrogen-like orbitals with usual values of the quantum numbers. 
It could be expected that such visualizations are readily available. Unfortunately, that is not the case. 
We illustrate the problems by showing several less fortunate earlier presentations 
in some scientifically most valuable works, and show more suitable visualizations
for those cases. The selected cases are not chosen to criticize the authors' approach. Rather,
we have taken these very important papers to underline the need for serious work
with graphical representations which this work attempts to be a part of. 

In this text we discuss the problems encountered when visualizing Rydberg states, review some 
existing presentations and propose 
guidelines for applications. The focus of this work are so called Stark states and coherent 
elliptic states of Rydberg atoms. In the sections on elliptic states we show some quite novel
geometric interpretations which should be very helpful in analysis of Rydberg state manipulations
in external fields.

Our demonstration examples involve only hydrogen-like states. However, the techniques and analysis
are  also relevant for Rydberg molecules and more complicated highly excited atomic and 
molecular systems.
 \ \ \ \  \\
  \ \ \ \  \\
   \ \ \ \  \\
  \ \ \ \  \\
   \ \ \ \  \\

  \ \ \ \  \\

\end{abstract}

\maketitle

%
%
%
\section{                   Introduction             \label{intro_section}        }
%
%
%
%
%
In theoretical physics it is sometimes assumed that once we have an analytical formula for
a certain quantity, the problem is solved, this quantity is known and 
for many purposes in principle understood. 
For Rydberg states (see e.g. Gallagher \cite{Gallagher_book}), whose analytic formulae are known since the first days of quantum mechanics
this definitely is not the case. Even though their formulae are well known, the structures can show
so much complexity of features that good visualization work is required. 
Ever since the first papers on the subject some graphical representations have
been included both in textbooks and in scientific papers.

The wavefunctions for Rydberg states are after all hydrogen-like wavefunctions and
the representation of hydrogen-like wavefunction appears in some form in many textbooks. 
They are also attractive for many popular science presentations. Recent offerings even include 
pictures similar to those discussed here. It might thus seem unnecessary to devote a special 
publication to such well known representations, but as we shall show, general
type of visualization of such states offers several challenges and
we will also discuss several previously published less fortunate cases. The main source of problems  
is that the application to large 
principal quantum number, i.e. weakly bound - or highly excited - states with many nodes, must
show structures with many details which might appear at varying scales. 
For large quantum number $n$ even the s-states, 
present some challenge.

In this paper we discuss generally the visualization of Rydberg states, 
with main focus on the so called Stark states and coherent elliptic states (CES), as well as
some related high $n$ states.
In section \ref{s_states_section} we discuss the seemingly straightforward visualization of 
probability densities or wave functions of the isotropic $s$-states and demonstrate the challenges 
present for highly excited Rydberg states and show possible solutions.

In section \ref{stark_section}  we show the problems of Stark state visualization. 
They have axial rotational symmetry, thus in principle two dimensional cuts by a plane
can be used, but it is demonstrated that a three-dimensional representation might be useful.
In section \ref{stark_unfortunate} we list and shortly discuss the first set of unfortunate representations
in scientifically important works.

The coherent elliptic states are presented in section \ref{Elliptic_section}. 
These states are truly three-dimensional and we discuss the method of visualization
using isosurfaces and their modifications, with cuts and transparency properties. 
In this context a new type of visualization is discussed, but it is also shown that similar
ones have been employed to related problems nearly 20 years ago.

Certain aspects of interpretation of coherent elliptic states are discussed in
 section \ref{interpretCES}. Here we again find some less fortunate visualization
 effects which propagate into future works. A classical picture is offered
 as an alternative intuitive justification, as well as a study of
 probability current density of the CES.

In the following section \ref{Crossed_fields_section} we discussed new applications
to a more general type of coherent superposed states and possible new insights
in a contemporary project on experimental studies of Rydberg atoms using 
electric and magnetic fields manipulations.

The visualization work reported in this paper is based on the package 
MATLAB by MathWorks. MATLAB is widely used in teaching mathematics and physics
and in many areas of research. Similar techniques are however
available in many different frameworks, some of them as free software.
Thus the general features of the visualizations presented are certainly not 
limited to MATLAB.
%
%
%
%
%
%
                           \section{Visualizing s-states                   \label{s_states_section}}
%
%
%
%
%
%
%
The s-states are radial functions only, thus a representation in terms of a single-variable plots should 
be sufficient to record the structure of the atomic state. There is still one step from knowing the
radial dependence to visualizing how the spherical object really appears, but for most people 
this mental step is rather easy. One can imagine for example a cut through the sphere and on
the cut to have displayed the map of the density. Every physicist has seen
through the education such representations of Earth or other nearly spherical objects.

Plotting a function of one variable should not present any problem - but for hydrogenic states
$\psi_{nlm}(\nvec{r} ) $ with large values of $n$ there appears immediately one problem - 
widely varying scale of magnitude from the smallest radial distances outwards.
%
%
%
%
\begin{figure}[t]
\begin{center}
\includegraphics[width=1\columnwidth]{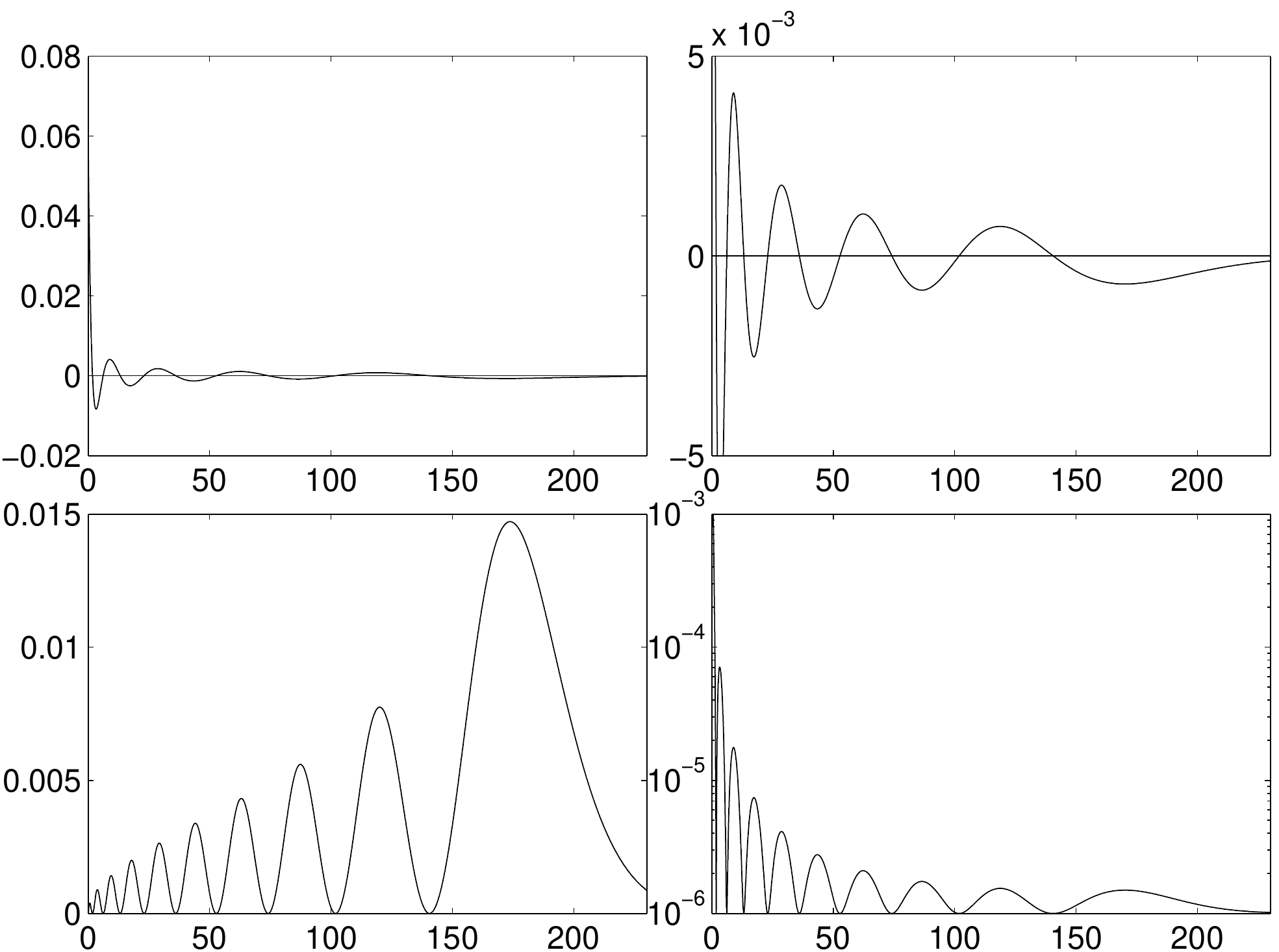}
\caption{  Plotting of radial wavefunctions and densities. Plot (a) shows the radial wavefunction plot for 
$n$=10, $l$=0. The plot is not useful, large r behavior is not represented. Plot (b) is the same with changed scale, which naturally eliminates a good representation of the small r region. Plot (c) is the well known plot showing the radial probability density. The success of this plot is a root of the discussed problems. Plot (d) shows a starting 
point for the discused solution: real probability density is plotted, but in the form of its logarithm, with cut-off for small values (large negative logarithm cut-off). This last plot illustrates that the logarithmic plotting might make it 
possible to construct good representations of the wavefunction (probability) structures for Rydberg states.  
\label{radial_fig}}
\end{center}
\end{figure}
%
%
In  figure \ref{radial_fig} we illustrate the origin of this problem, here on the example of the 
radial function of an s-state $n$=10, $l$=0,  a relatively low $n$.

The figure first shows a  plot of radial wavefunction in section (a). Clearly the plot is not very useful for 
displaying the large r behavior on the scale of values relevant for small r. 
One straightforward method is simply to change scale, which naturally eliminates a good 
representation of the small r region, but that does not matter since one knows that the states
contain the exponential decay modified by a combination of polynomials and thus is simply a monotonous 
steep fall with r until it enters the picture. It should however be kept in mind
that there is a small radial region where the wavefunction peaks enormously for large $n$.
Generally, Rydberg states have 
interesting structures and also majority of their probability density in regions where the actual
probability density is low. Maximal values of the density appear close to the nucleus, but the total
relative contribution of this is close to negligible. 

Plot (c) is the well known type of plot often used in textbooks showing the radial probability density. 
The success of this plot is a root of problems and misunderstandings discussed below. It is so
popular that sometimes "plot of radial density" is understood as this type of plot. One can misplace
"radial dependence of density" - a well defined term with "radial density" - a term which does not
have a unique definition, but is often understood as the shown picture (c).
Textbook definition might be: {\it The radial probability density for the hydrogen ground state is obtained by multiplying the square of the wavefunction by a spherical shell volume element}. 
We see that this picture completely distorts the relative importances. The maximum values are
unfortunately eliminated, so this representation conceals the fact of extremely
high relative densities at the nucleus and exaggerates the values at large distances. 

Plot (d) shows a starting 
point for one of our proposed solution: actual probability density is plotted, but in the form of its logarithm, with cut-off for small values (large negative logarithm cut-off). This last plot illustrates that the logarithmic plotting might make it 
possible to construct good representations of the wavefunction (probability) structures for Rydberg states. It is a well known 
technique to show multi-scale dependencies. 

The problems discussed here for states with maximum symmetry become necessarily
much more complex for states with less symmetry and the visualization task
becomes much more difficult. In the next section we shall show how this has been a cause of more or less
confusing information in the literature.

 When moving to general $n$, $l$, $m$, the densities of states for all of these are rotationally
symmetric with respect to z-axis, since they are eigenstates of $L_z$. Thus for any $n$, $l$, $m$, state a plot in terms of 2 variables is sufficient for all $n$, $l$, $m$, states.
\begin{center}
\begin{tabular}{  l  l  l}
$n00$             &  \ \ \ & simple radial plot                   \\
$nlm$              &  \ \ \ & plot of a plane cut (x, z) - 2 variable plot \\
superposition    &  \ \ \ & depends on x, y, z, 3-dimensional  \\
\end{tabular}\\
\end{center}
It is thus clear that for a unified presentation, we need some methods
to present the $| \Psi(x,y,z) |^2$
%
%
%
%
%
%
%
%
                    \section{Stark states                  \label{stark_section}}
%
%
%
%
%
%
Already in his 1926 paper Pauli (\cite{Pauli}, note the English translation 
available)  described the solutions of 
Schr\"odinger equation for electrons in electric and magnetic field, and he
has built on already published earlier works, so some of the aspects 
discussed here have been part of the general knowledge
from the beginning of the Quantum mechanics. The states in
electric field only, without magnetic field,
are important for understanding of the Stark effect and it has become usual
to refer to these states as Stark states. Instead of the quantum numbers $n$, $l$ and $m$ the 
Stark states are characterized by quantum numbers $n$, $k$ and $m$, i.e. 
the angular momentum quantum number $l$ is replaced by the Stark quantum number $k$. 
Sometimes one is using 
the parabolic quantum numbers $n_1$ and $n_2$ (e.g. Gallagher's book  \cite{Gallagher_book}), 
but the notation with quantum number $k$ mentioned above is 
more symmetric and will be of particular advantage in section 
\ref{Crossed_fields_section} on crossed electric and magnetic fields. 
 \newcommand{\flatleng}{8cm}
\begin{figure}[h]
\begin{center}
\begin{tabular}{ll}
\includegraphics[width=\flatleng] {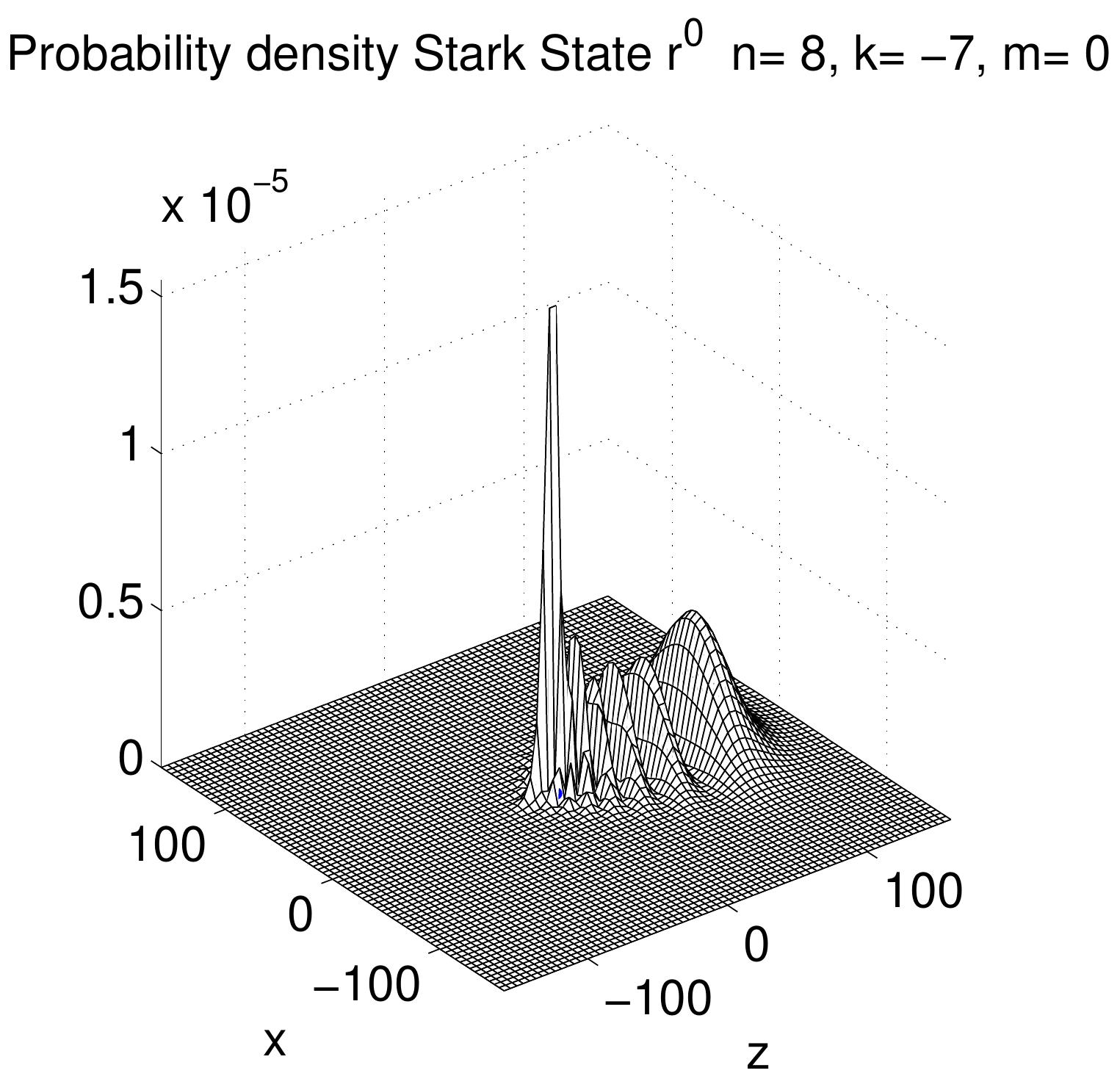} &
\includegraphics[width=\flatleng] {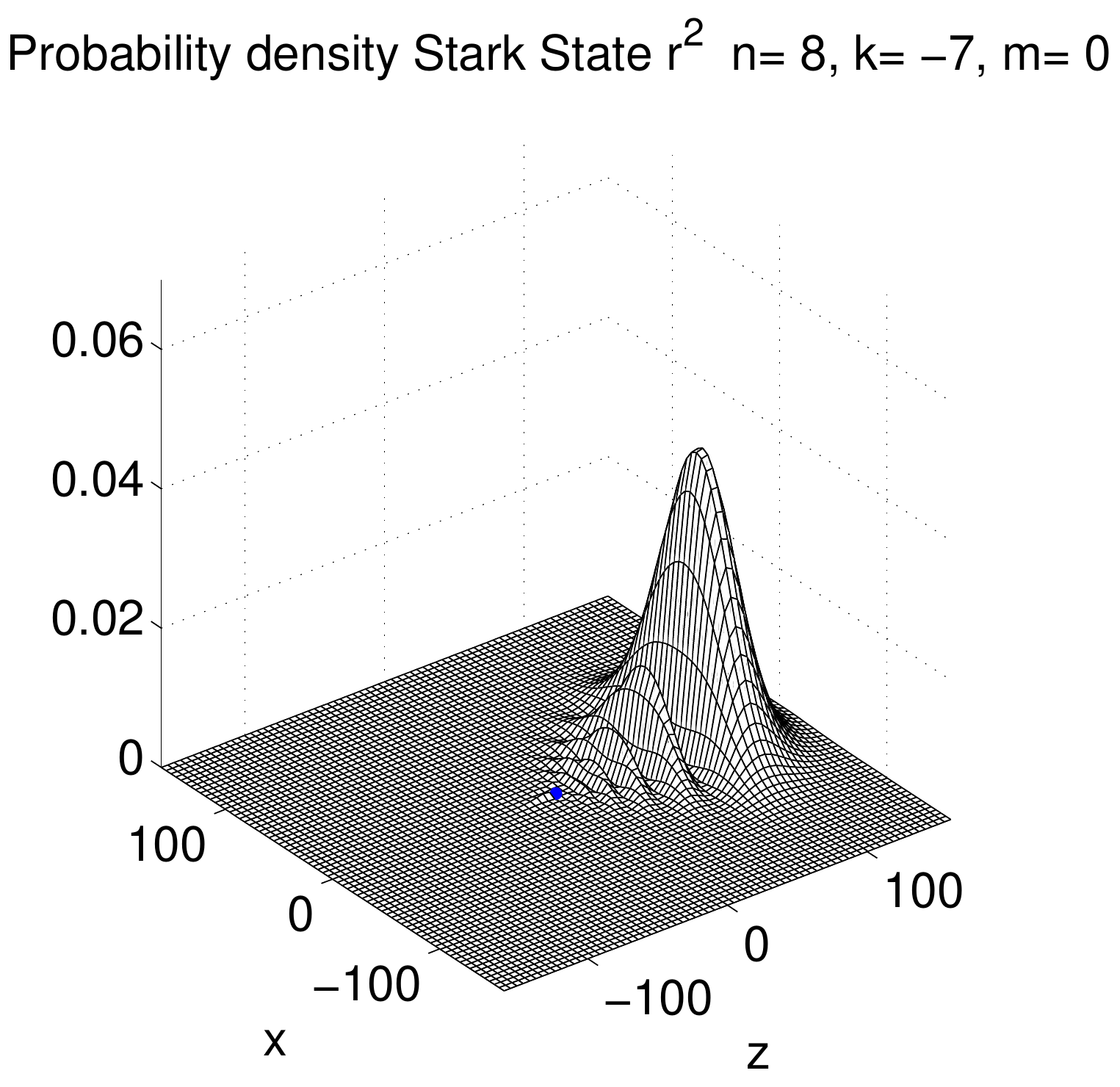} \\
\includegraphics[width=\flatleng] {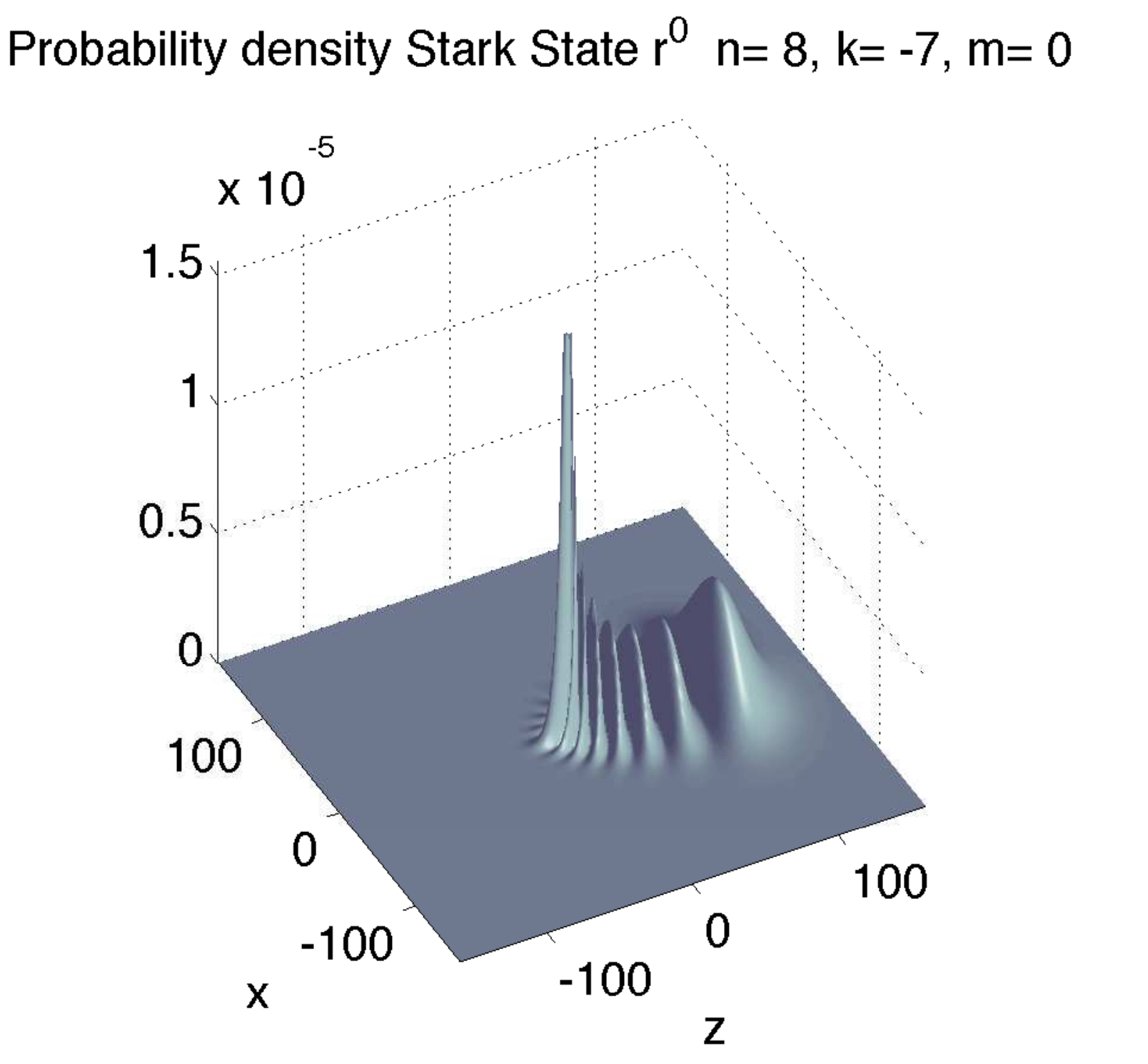} &
\includegraphics[width=\flatleng] {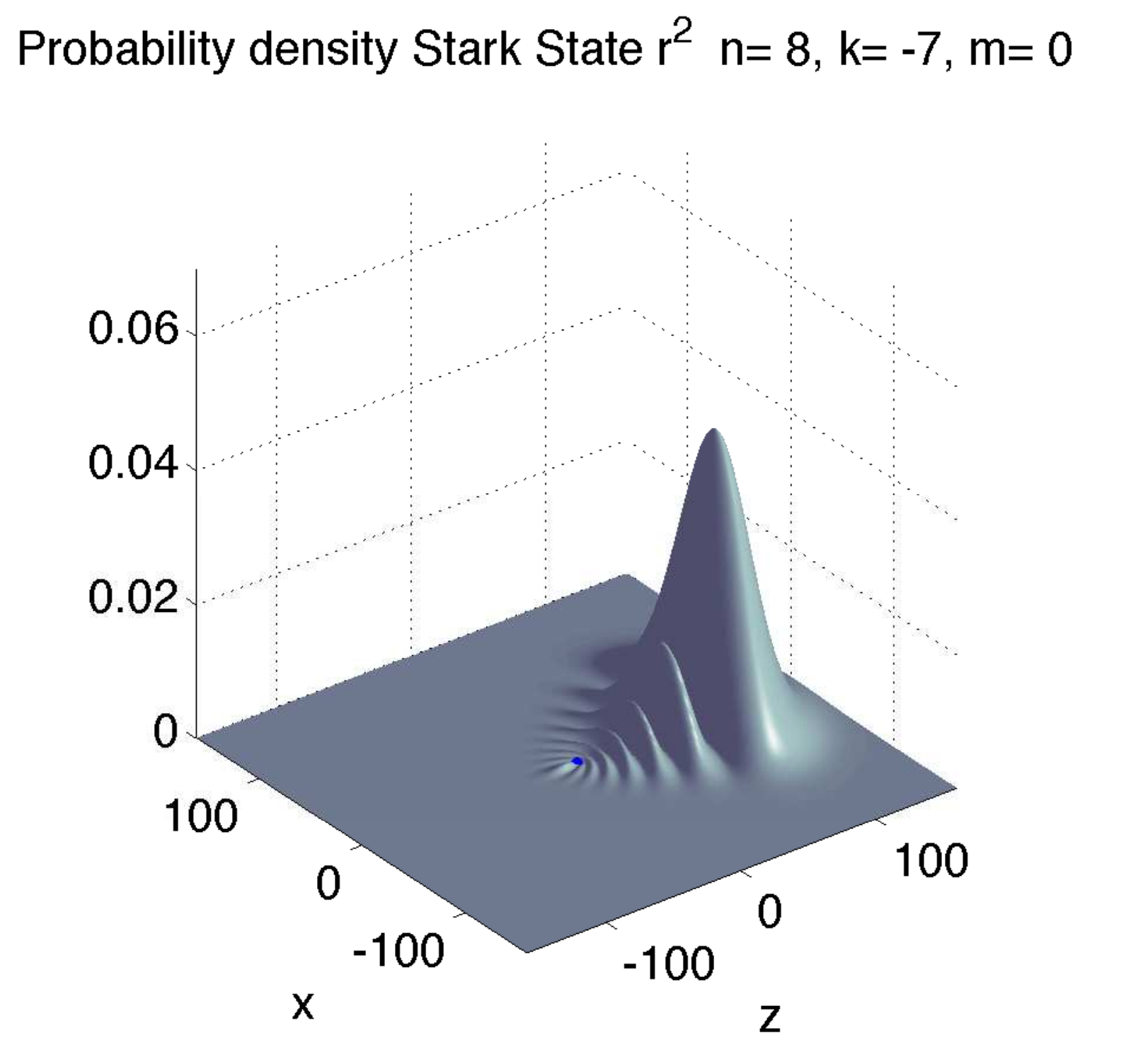}  \\
\end{tabular}
\caption{Landscape pictures of Stark states. The top row is
using a type of appearance usual in the 1980's. (a) is constructed in similar way
as in the s-states case: a cut-off is introduced. (b) The same data modified
to avoid the nuclear peak problem - multiplied by $r^2$. In the second row,
the same plots with a more modern appearance, the polygons are replaced by 
continuous shading. 
 \label{landscape-stark}    } 
\end{center}
\end{figure}

While the
usual spherical states are eigenstates of $L^2$ the Stark states are eigenstates of the coordinate $z$,
or in some sense more precisely the Runge-Lenz operator. For the purpose of the present discussion 
we limit ourselves only to the summary of Stark state properties. 

With this usual definition the Stark states have cylindrical symmetry
with z-axis as the symmetry axis. This property they share with the
usual set of spherical eigenstates which are also characterized by the
magnetic quantum number $m$. With cylindrical symmetry the probability
densities are functions of two variables. As in the s-states case discussed
in section \ref{s_states_section} their
display can be accomplished in a lower-dimensional space, in this case
as a function of two variables. The problems will be very similar
in this case, but amplified by the broader variation of possible display methods.

Probably the oldest way to display function of two variables are the maps
showing the heights using equal height curves (contour maps) or color coded
representations. They are most precise and suitable for displaying shapes
of the different areas, since they are seen "from above" and preserve the 
x-axis and y-axis (in this case the cylindrical $\rho$ and $z$ along the 
electric field direction). Another possibility which became very
popular in the first powerful years of computer graphics (say the 1980's)
is a quasi-three-dimensional plot where one generates the landscape
plot of the function of two variables. 

One would usually choose not to plot only the defined positive $\rho$, but
would rather show a cut through the density function e.g. along the (x-z)-plane,
showing an reflection symmetric pattern of the density.

The landscape plots examples are shown in figure landscape-stark. If we start plotting
the $| \Psi(x,y,z) |^2$ we quickly encounter the same problem of the very high 
nuclear maximum - and the same methods as discussed in section \ref{s_states_section}
apply here. But now we can confuse the issues even more. We can choose to plot a sort
of radial density $2 \pi \rho | \Psi(x,y,z) |^2$ in analogy with the spherical definition.
That might not help much, the nuclear peak in the landscape map still remains very 
prominent. Thus during the years many people have chosen to plot $ r^2 | \Psi(x,y,z) |^2$
also in this case. 

This choice has the advantage to display all the regions, but it has many disadvantages. 
First of all, one must not forget to mention what is actually plotted. Many people
through the years and even most recently forget to do that. Second, the shapes 
might be quite distorted relatively, it might be necessary to choose varying viewpoints
for different functions and so on.

One unfortunate accident is that a display of these problems can be
quite  exactly illustrated by Figure 6.1 of the classic work, 
Gallagher's book Rydberg Atoms (\cite{Gallagher_book}, page 73). 
The figure appears to be a type of reprinted information so that the 
author can is not fully responsible for the missing and slightly
misleading information (the caption says "Charge distribution for H for parabolic eigenstates",
but what is plotted are the modified distributions, i.e. multiplied
by a power of $r$, as far as we can see by the factor $r^2$).
The main point of the figure caption is the
illustration of the induced dipole moments - and that point is definitely correct.

To complement this excellent classic book by Gallagher
we bring here our version of  its figure 6.1 on page 73
in the present figure \ref{flatfig_gallagher}.
\clearpage
 \newcommand{\figdim}{5cm}
\begin{figure}[h]
\begin{center}
\begin{tabular}{ll}
\includegraphics[width=\figdim] {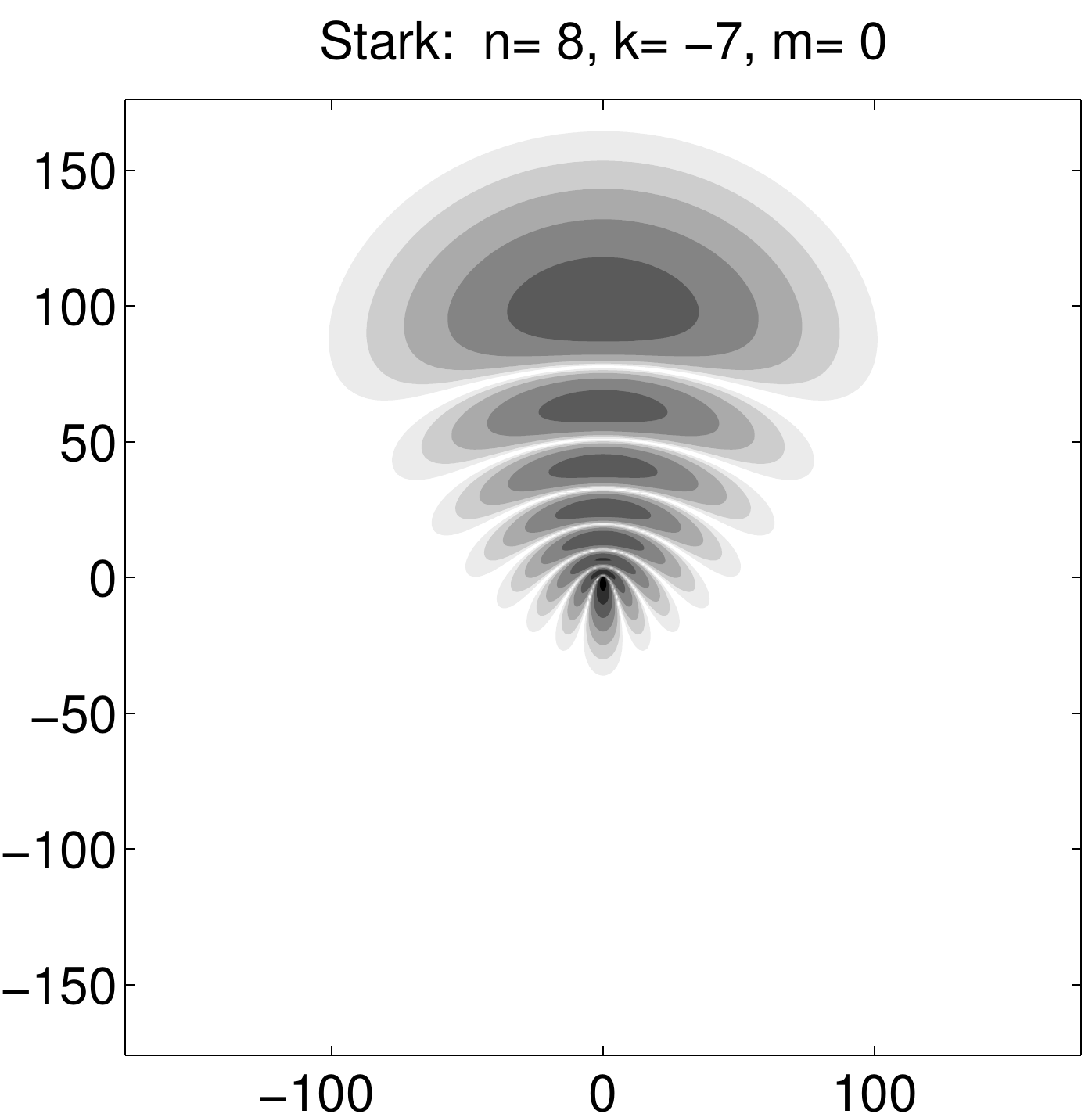} &
\includegraphics[width=\figdim] {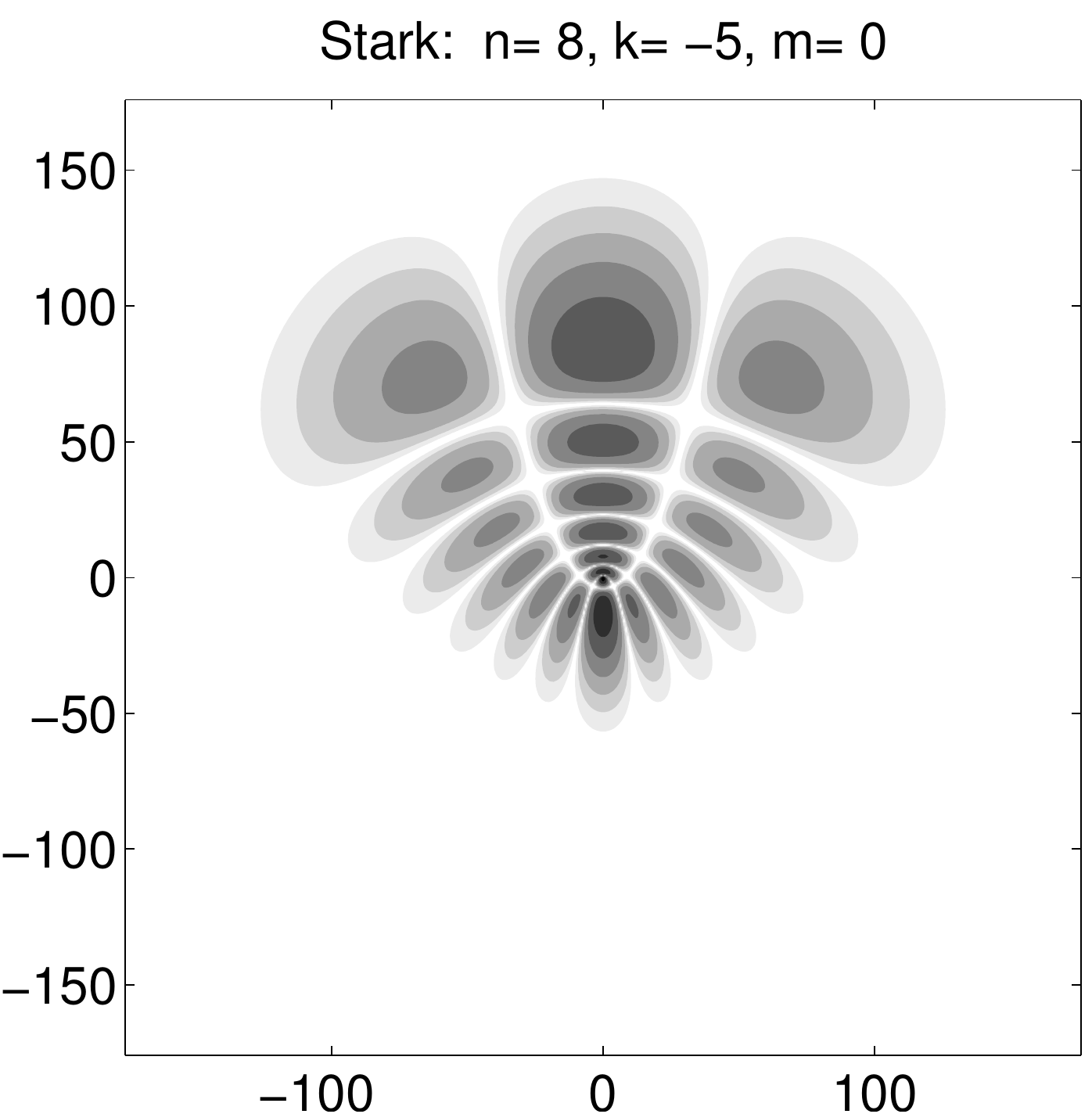} \\
\includegraphics[width=\figdim] {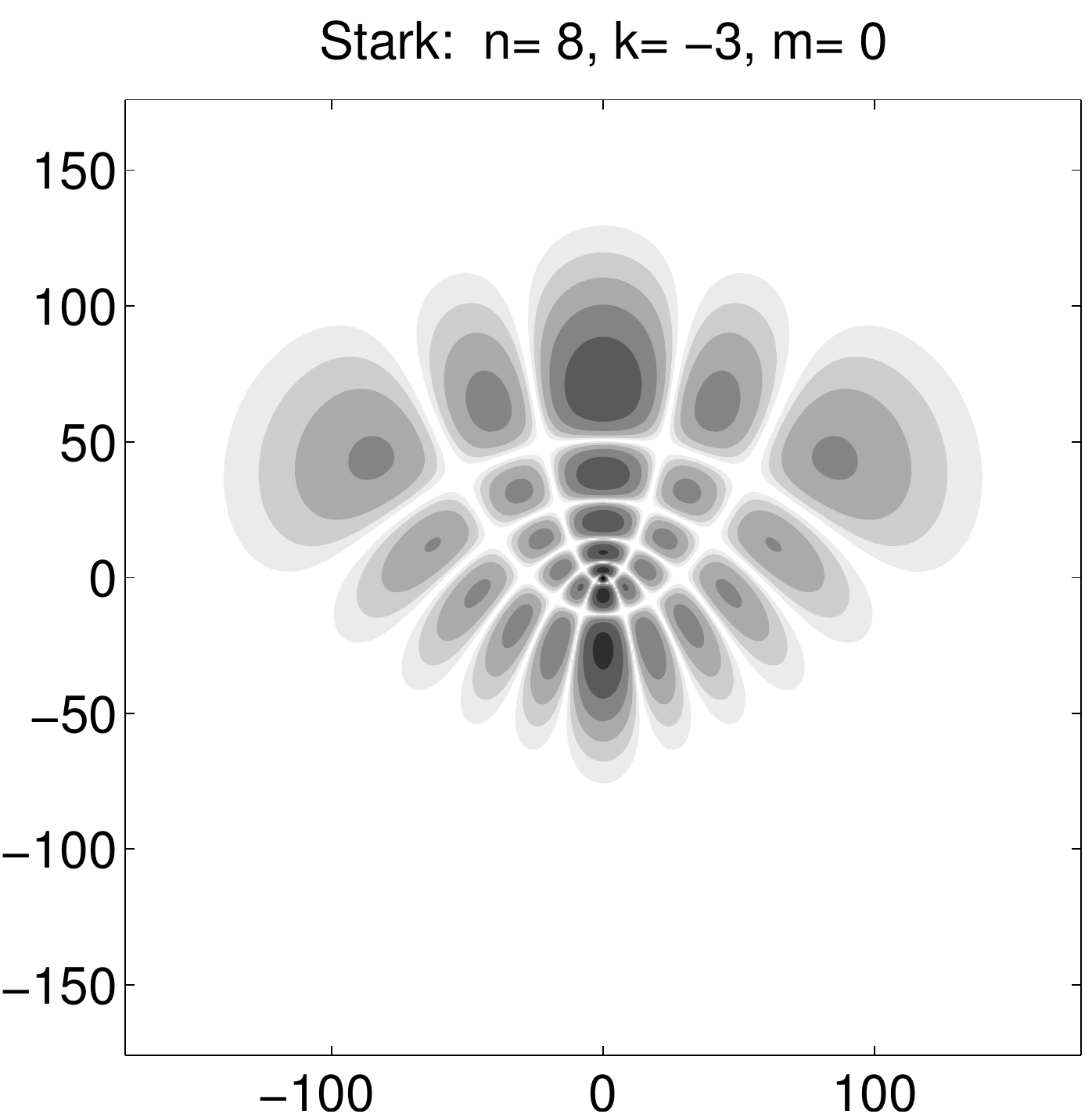} &
\includegraphics[width=\figdim] {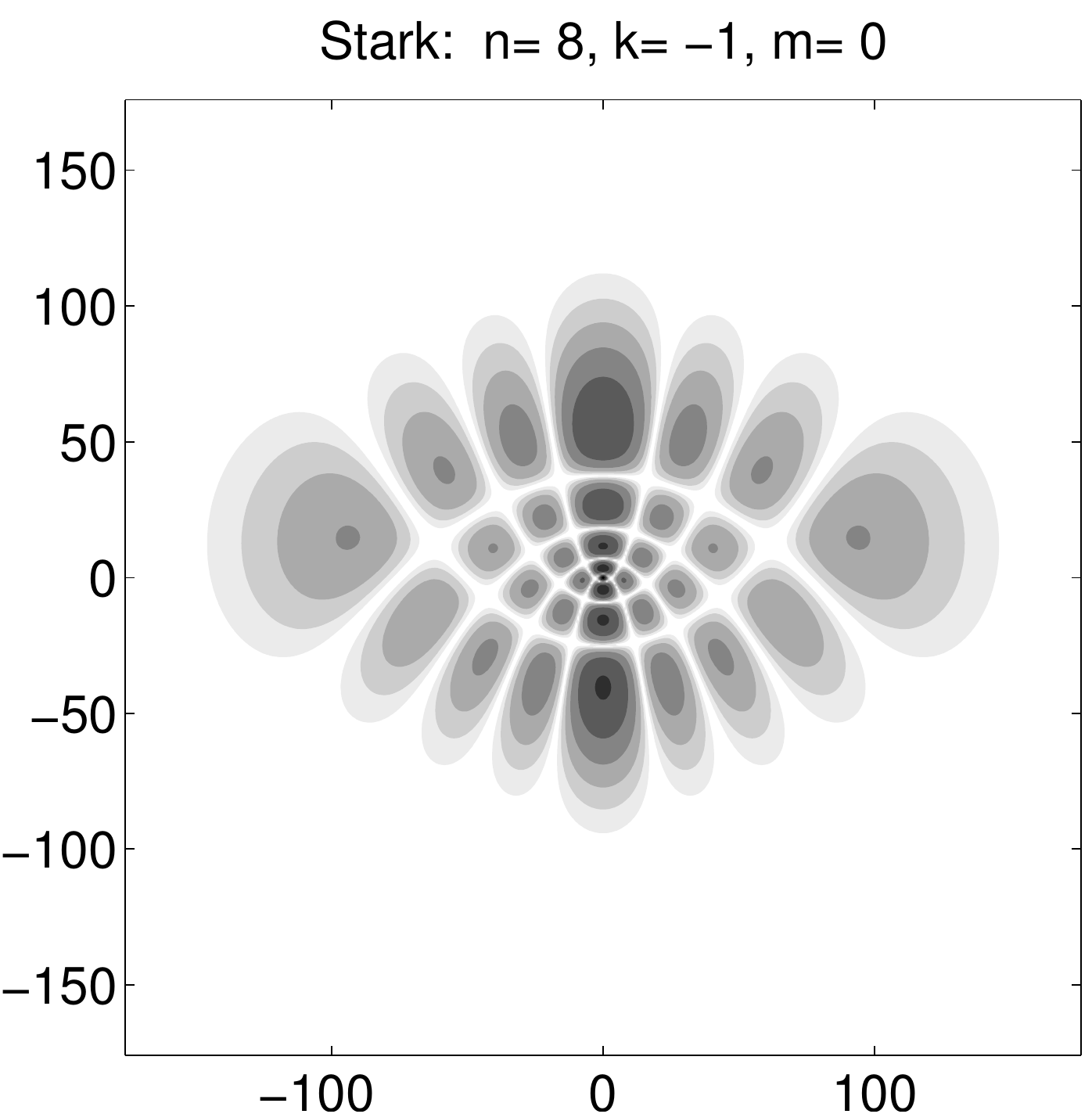}  \\
\includegraphics[width=\figdim] {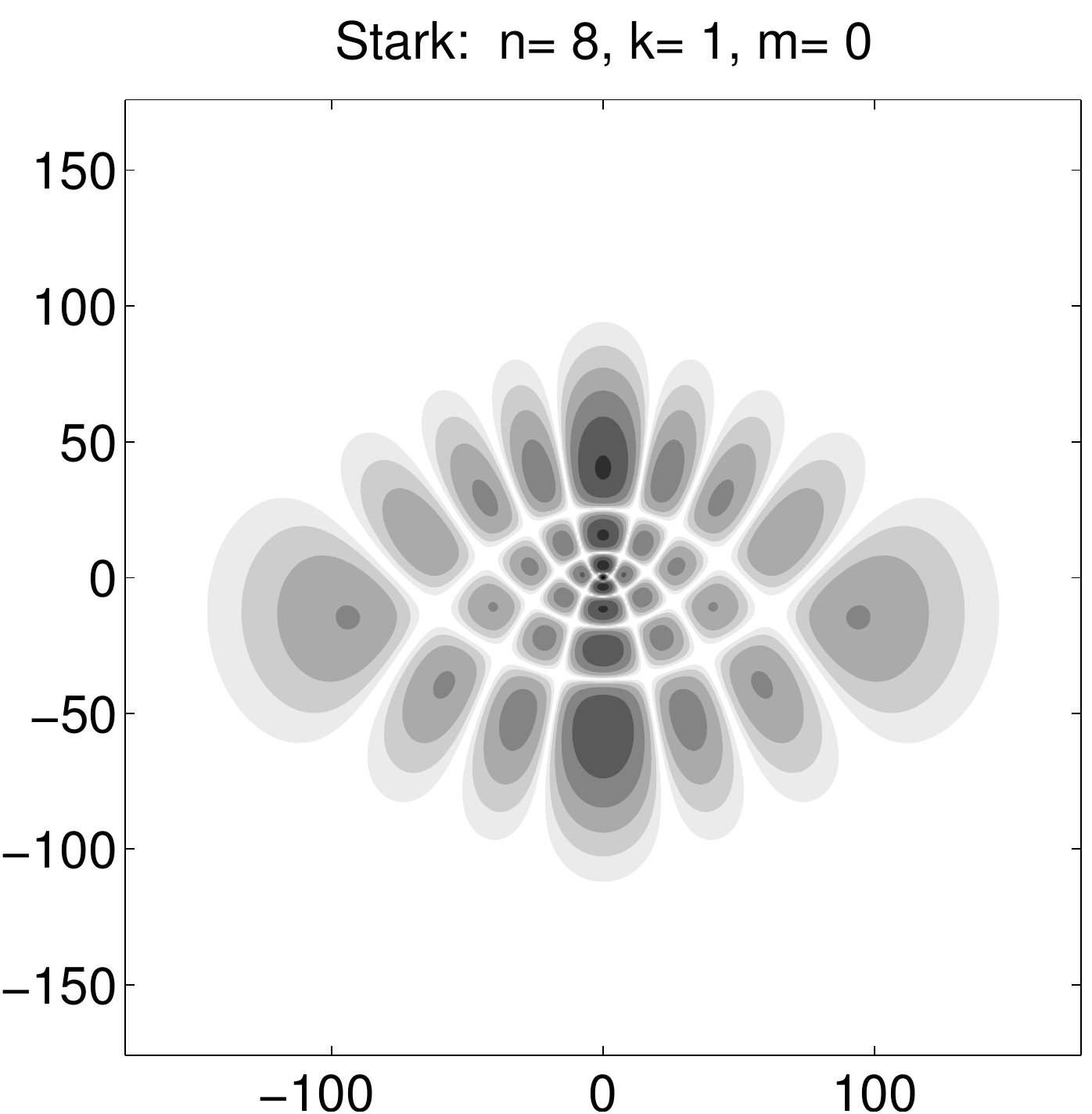}&
\includegraphics[width=\figdim] {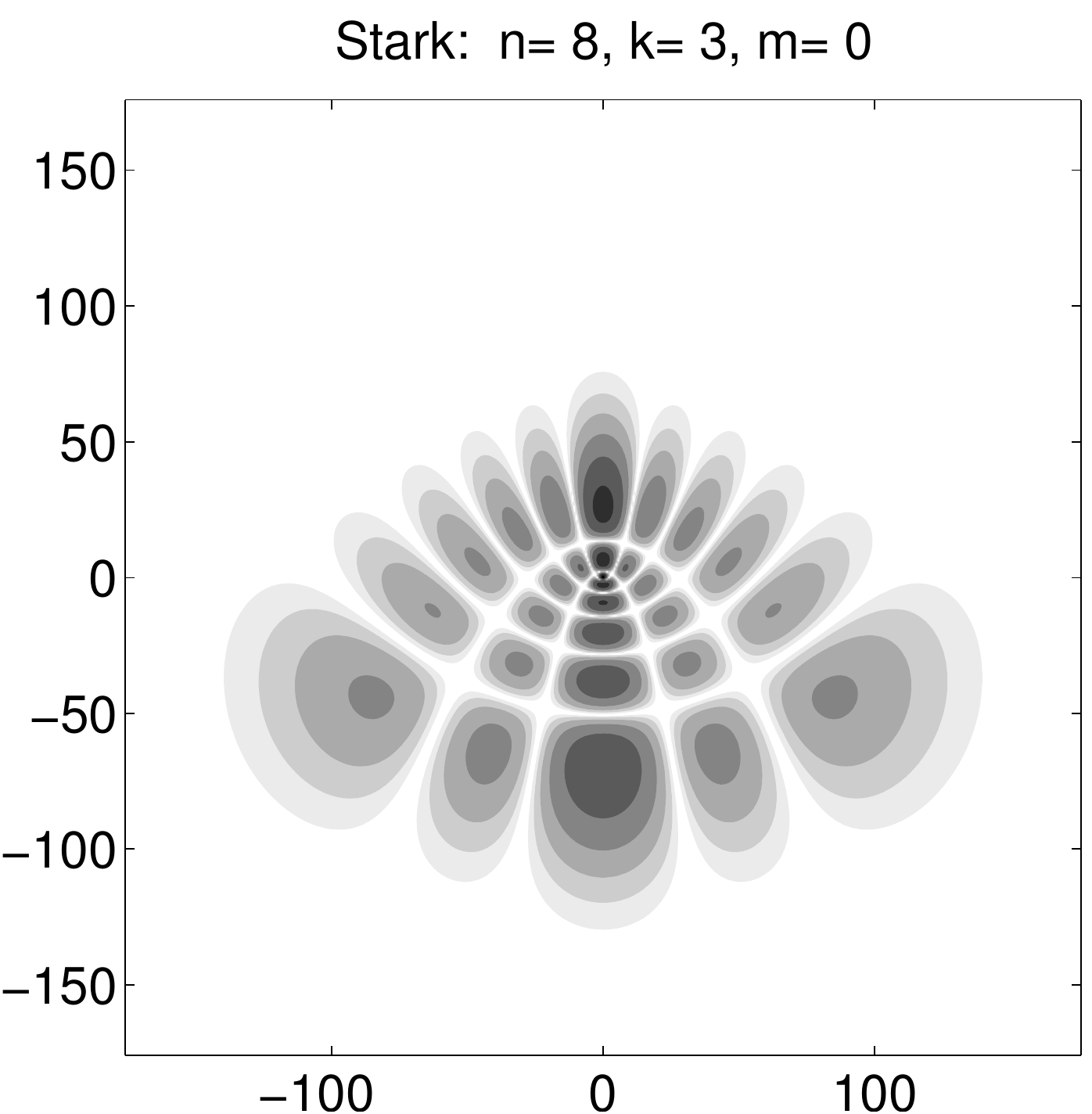}\\
\includegraphics[width=\figdim] {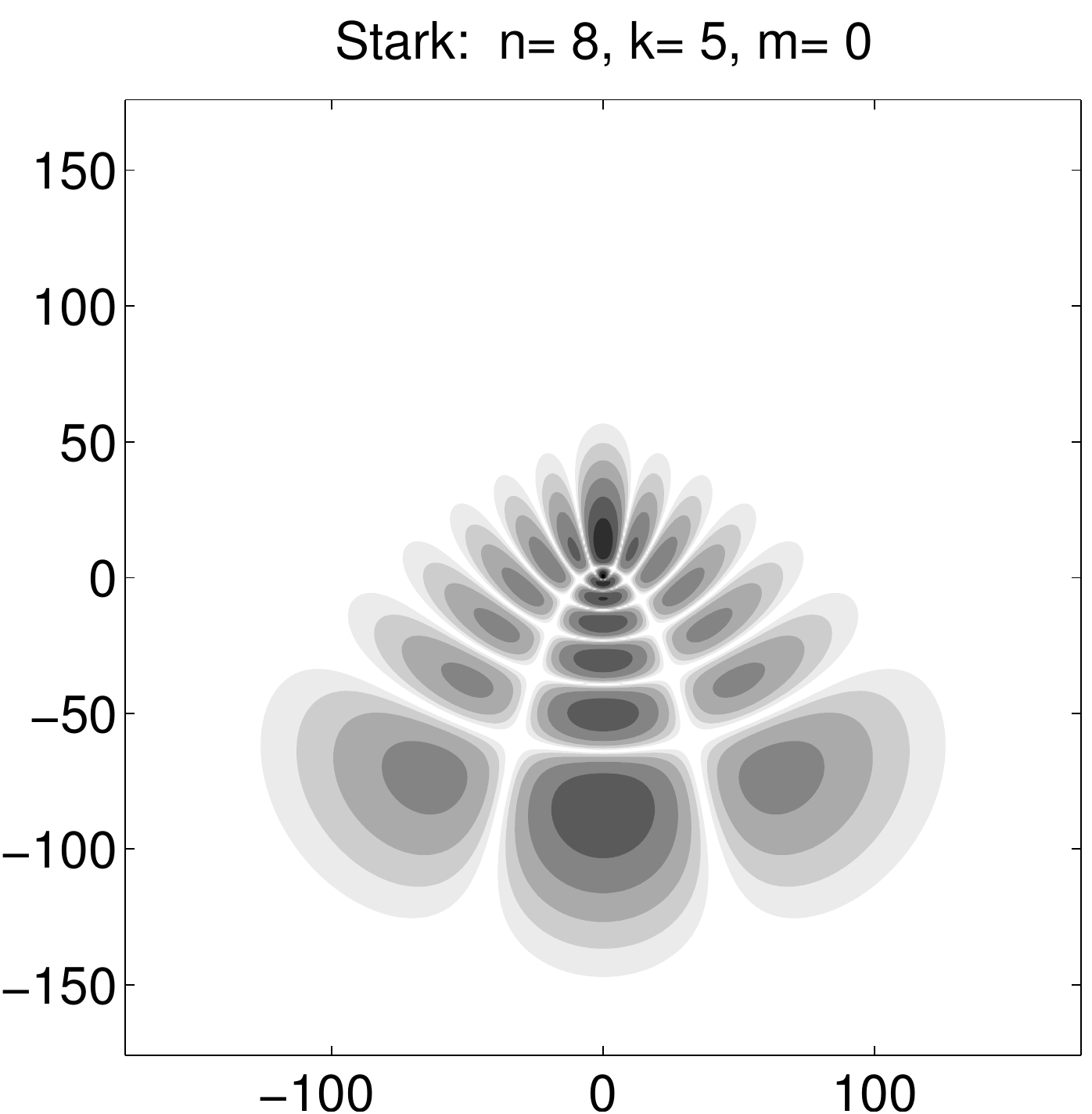}&
\includegraphics[width=\figdim] {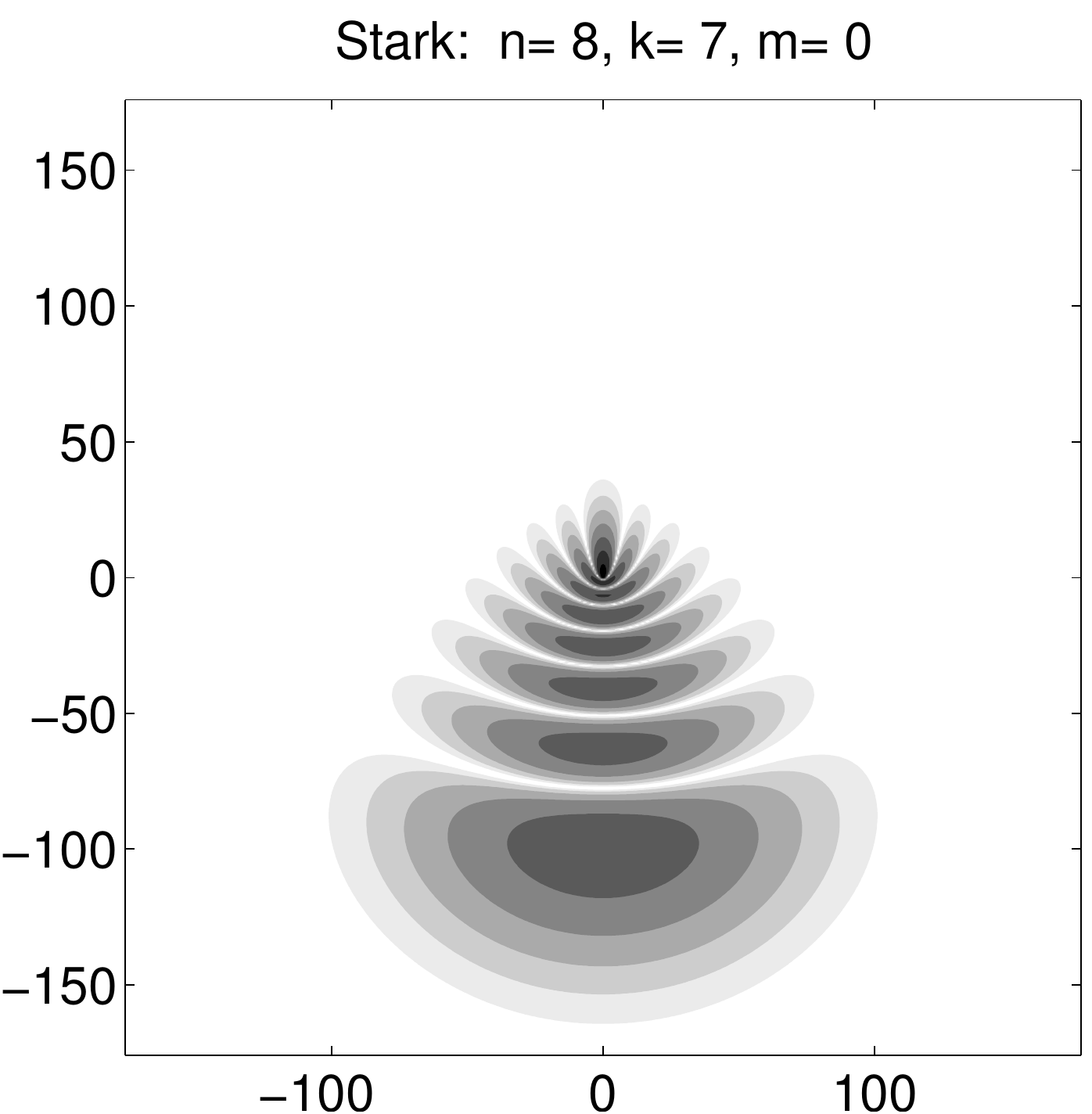}\\
\end{tabular}
\caption{Replacement for Gallagher  Rydberg atoms figure 6.1.
This type of plot is to be preferred. The density is visualized by plotting a contour 
plot of the logarithm of the density, as used also in the one-dimensional case of 
radial density in figure \ref{radial_fig}
\label{flatfig_gallagher} } 
\end{center}
\end{figure}
\clearpage
%
%
In order to make the content of figure \ref{flatfig_gallagher}, i.e. the 
maps of probability density (which use logarithmic representation, each new shade denotes an
 \newcommand{\figthree}{7cm}
\begin{figure}[h]
\begin{center}
\begin{tabular}{cc}
\includegraphics[width=\figthree]{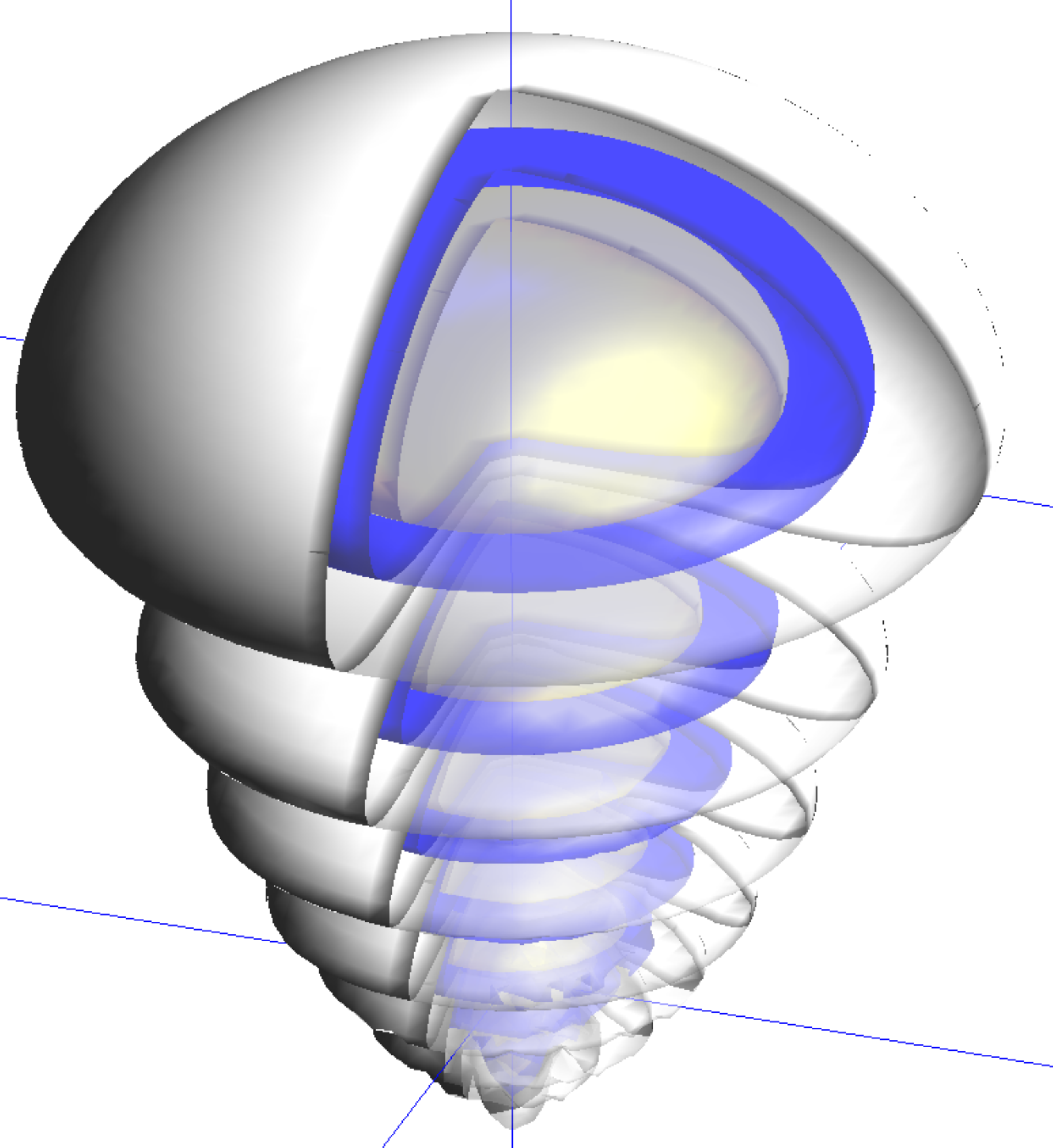}  &
\includegraphics[width=\figthree] {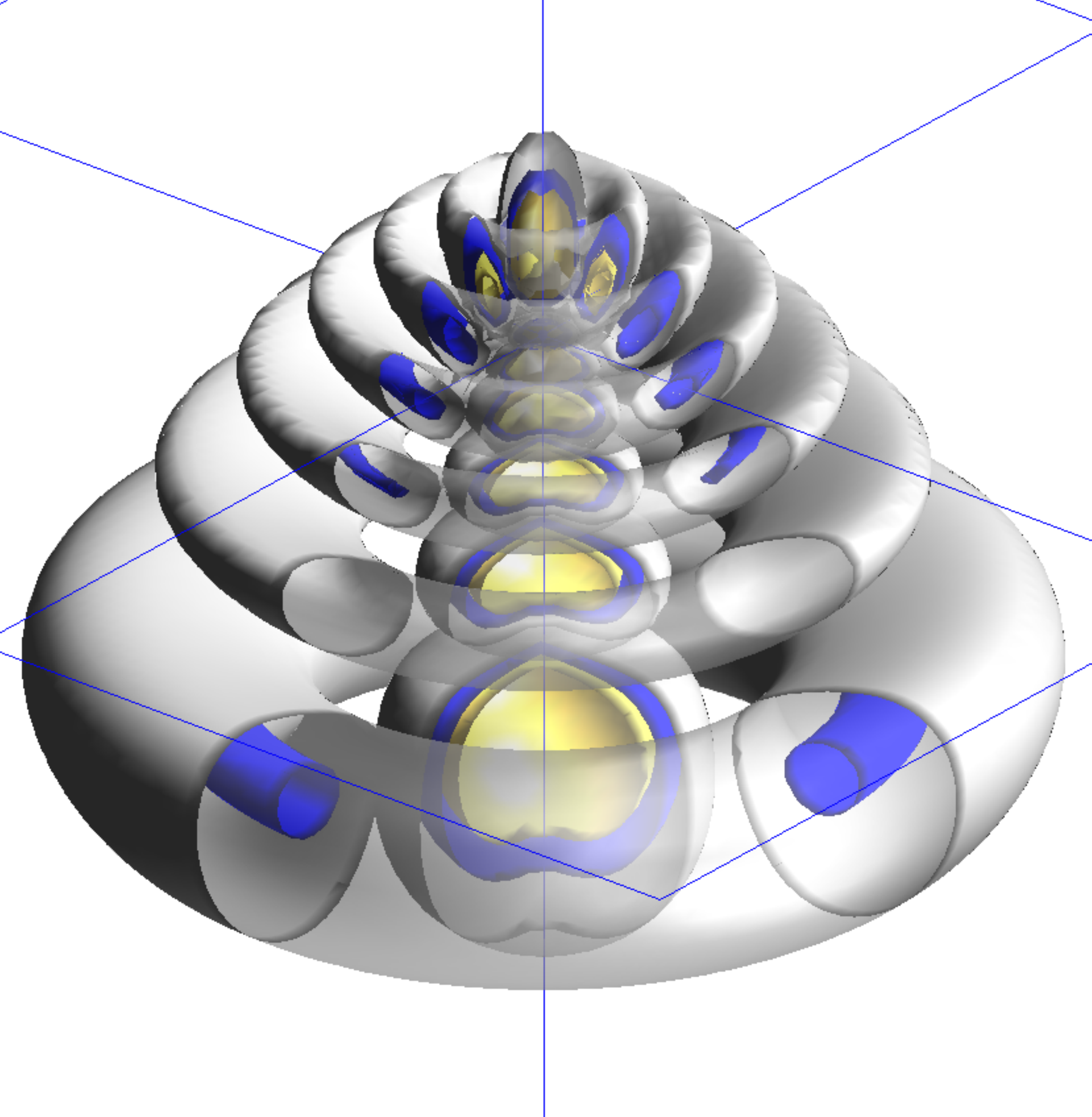} \\
$n=8\ \ k= -7 \ \ m=0 $  &
$n=8\ \ k= 5 \ \ m=0 $  \\
$ \ \ $  &
$ \ \  $  \\
\includegraphics[width=\figthree]{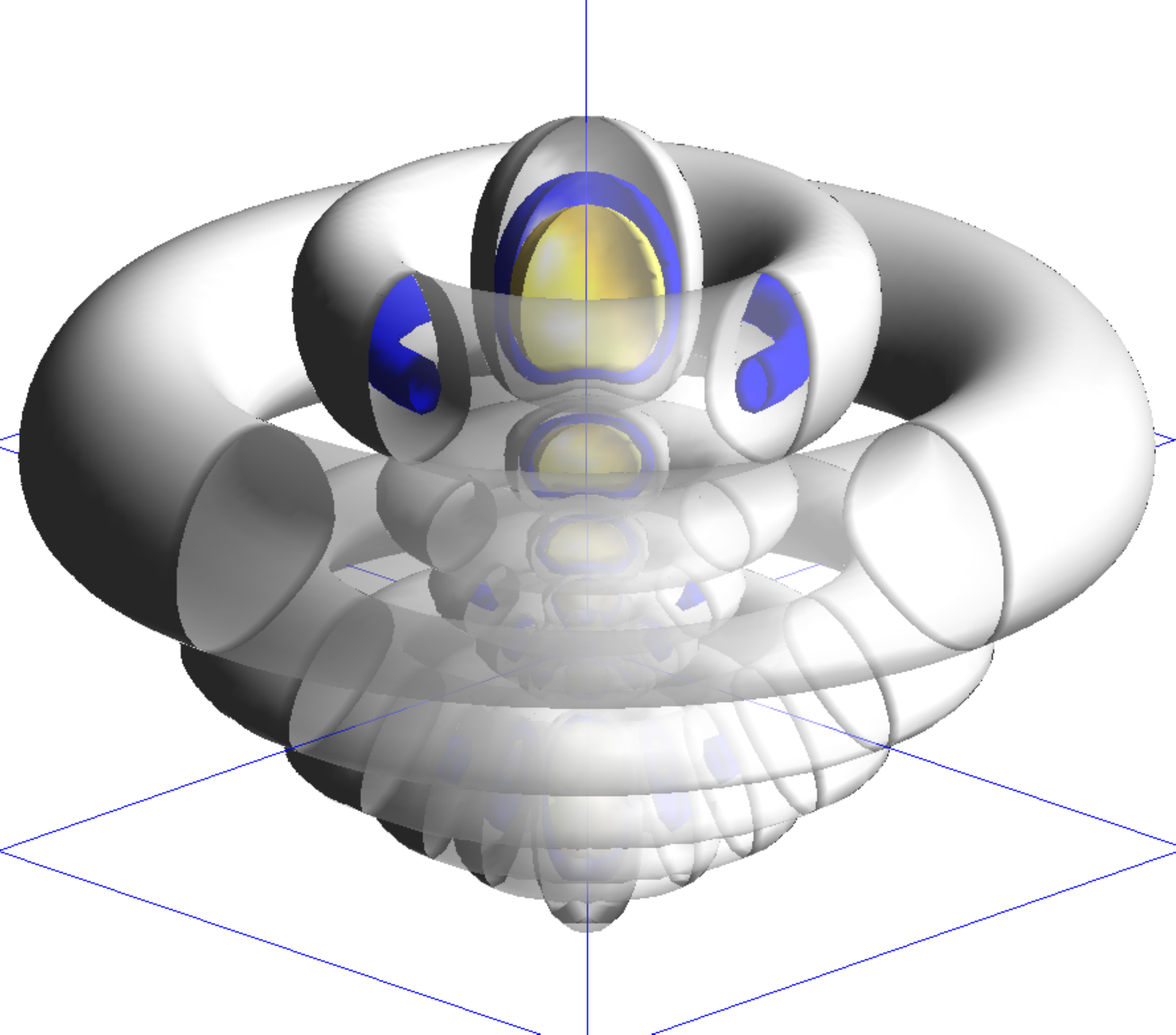}  &
\includegraphics[width=\figthree]{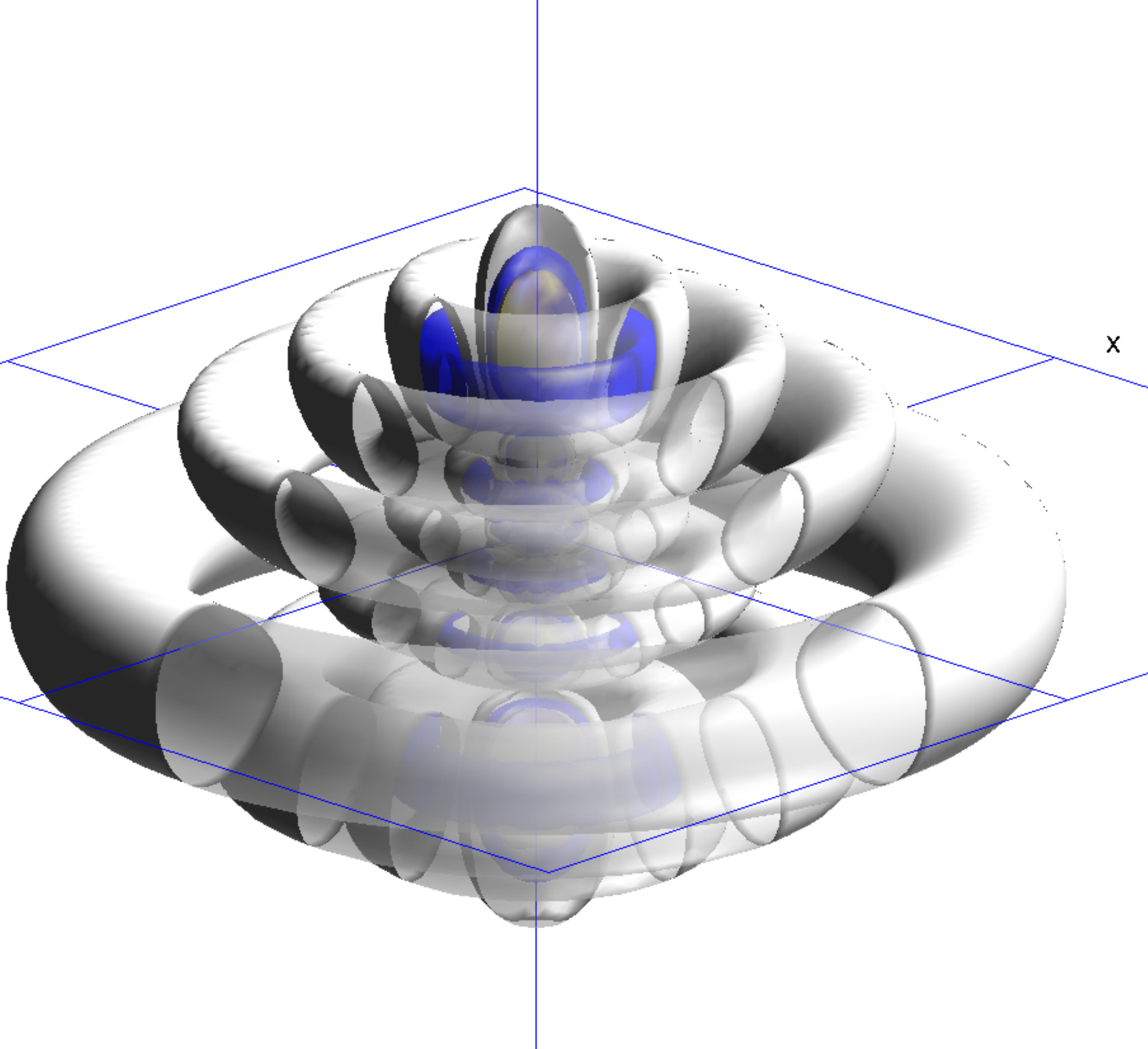}   \\
$n=8\ \ k= -3 \ \ m=0 $  &
$n=8\ \ k= 1 \ \ m=0 $  \\
\end{tabular}
\caption{The figures  $n= 8$, $|k|=1$ to 7, $m= 0$
are plotted. we take $k=-7$, $k=5$, $k=-3$ and  $k=1$  to show the various features
 \label{threedim_gallagher} }
\end{center}
\end{figure}
increase by about a factor $e$) more intuitive, we show a different three-dimensional representation of the
shapes in figure  \ref{threedim_gallagher}. In this figure we repeat only the four differing
shapes in the table of eight, alternating the positive and negative values from 7 to 1.

\begin{table}[htb]
\centering
\begin{tabular}{|c|c|c|c|c|c|   }      \hline
 \multicolumn{6}{|c|}{n=5}\\ \hline
 m     & \multicolumn{5}{|c|}{l  } \\ \hline
-4&    4&   &	&    &    \\  \hline
-3&    4&  3&	&    &    \\   \hline
-2&    4&  3&  2&    &    \\    \hline
-1&    4&  3&  2&   1&    \\ \hline
 0&    4&  3&  2&   1&   0\\	   \hline
 1&    4&  3&  2&   1&    \\	   \hline
 2&    4&  3&  2&    &    \\ \hline		   
 3&    4&  3&	&    &    \\ \hline
 4&    4&   &	&    &    \\ \hline
\end{tabular}
\ \ \ \ \ \ \ \ \   \   
\begin{tabular}{|c|c|c|c|c|c|c|c|c|c|  }      \hline
 \multicolumn{10}{|c|}{n=5}\\ \hline
 m     & \multicolumn{9}{|c|}{k  } \\ \hline
-4 &\ 	&\   &\   &\   &  0 &\   &\   &\   &\	 \\  \hline
-3 &\   &\   &\   &-1  &\   & 1  &\   &\   &\	 \\   \hline
-2 &\   &\   &-2  &\   &  0 &\   & 2  &\   &\	 \\    \hline
-1 &\   &-3  &\   &-1  &\   & 1  &\   & 3  &\	 \\ \hline
 0 & -4 &\   &-2  &\   &  0 &\   & 2  &\   &  4  \\	 \hline
 1 &\   &-3  &\   &-1  &\   & 1  &\   & 3  &\	 \\	 \hline
 2 &\   &\   &-2  &\   &  0 &\   & 2  &\   &\	 \\ \hline		
 3 &\   &\   &\   &-1  &\   & 1  &\   &\   &\	  \\ \hline
 4 &\   &\   &\   &\   &  0 &\   &\   &\   &\	 \\ \hline
\end{tabular}
\caption[Stark and Spherical Basis States, $n=5$]{Stark states combinations of hydrogen atom for $n=5$ to the right, compared to
the states of spherical basis arranged in similar manner to the left\label{compare_nlm_nkm_5}}
\end{table}

\begin{table}[htb]
\centering
\begin{tabular}{|c|c|c|c|c|   }      \hline
 \multicolumn{5}{|c|}{n=4}\\ \hline
 m     & \multicolumn{4}{|c|}{l  } \\ \hline
-3&    3&   &	 &    \\   \hline
-2&    3&  2&	 &    \\    \hline
-1&    3&  2&	1&    \\ \hline
 0&    3&  2&	1&   0\\       \hline
 1&    3&  2&	1&    \\       \hline
 2&    3&  2&	 &    \\ \hline 	       
 3&    3&   &	 &    \\ \hline
\end{tabular}
\ \ \ \ \ \ \ \ \   \   
\begin{tabular}{|c|c|c|c|c|c|c|c|  }      \hline
 \multicolumn{8}{|c|}{n=4}\\ \hline
 m     & \multicolumn{7}{|c|}{k  } \\ \hline
-3  &\   &\   &    &  0	&\   &\   &\    \\   \hline
-2  &\   &\   &-1  &\	& 1  &\   &\   \\    \hline
-1  &\   &-2  &\   &  0 &\   & 2  &\    \\ \hline
 0  &-3  &\   &-1  &\	& 1  &\   & 3    \\	\hline
 1  &\   &-2  &\   &  0 &\   & 2  &\   \\	\hline
 2  &\   &\   &-1  &\	& 1  &\   &\     \\ \hline	       
 3  &\   &\   &    &  0	&\   &\   &\     \\ \hline
\end{tabular}
\caption[Stark and Spherical Basis States, n=4]{Stark states combinations of hydrogen atom for n=4 to the right, compared to
the states of spherical basis arranged in similar manner to the left\label{compare_nlm_nkm_4}}
\end{table}
%
%
%
%

This type of figures will be discussed in more detail below, here we just mention that
these are iso-surface plots, three-dimensional generalisation of the contour plots. To show the inner structure
of the distributions, part of the isosurfaces are made transparent. Here only three values of density are used.
The densities are normalized by factors of $n^6$-type and isosurfaces with values 0.5, 2 and 4 are used. 
No logarithmic scale is used in this case.
Note that all the states shown in figures  \ref{flatfig_gallagher} and \ref{threedim_gallagher}
are  $m=0$ states, it means they are 
unique and they are real functions, it means the numerous "tubes" do not enclose
any currents. Usually such tubes appear for $|m|>0$ states where 
they 'enclose'  an "azimuthal" current, since for $|m|>0$
the wavefunctions contain a factor $\exp (i m \phi)$ as in the spherical harmonics.
It should however also be mentioned that quite rich structures of somewhat
similar type are also obtained for high $n$ and $l$ with $m$=0. 

 \newcommand{\figdefdim}{4.6cm}
\begin{figure}[h]
\begin{center}
\begin{tabular}{lll}
\includegraphics[width=\figdefdim] {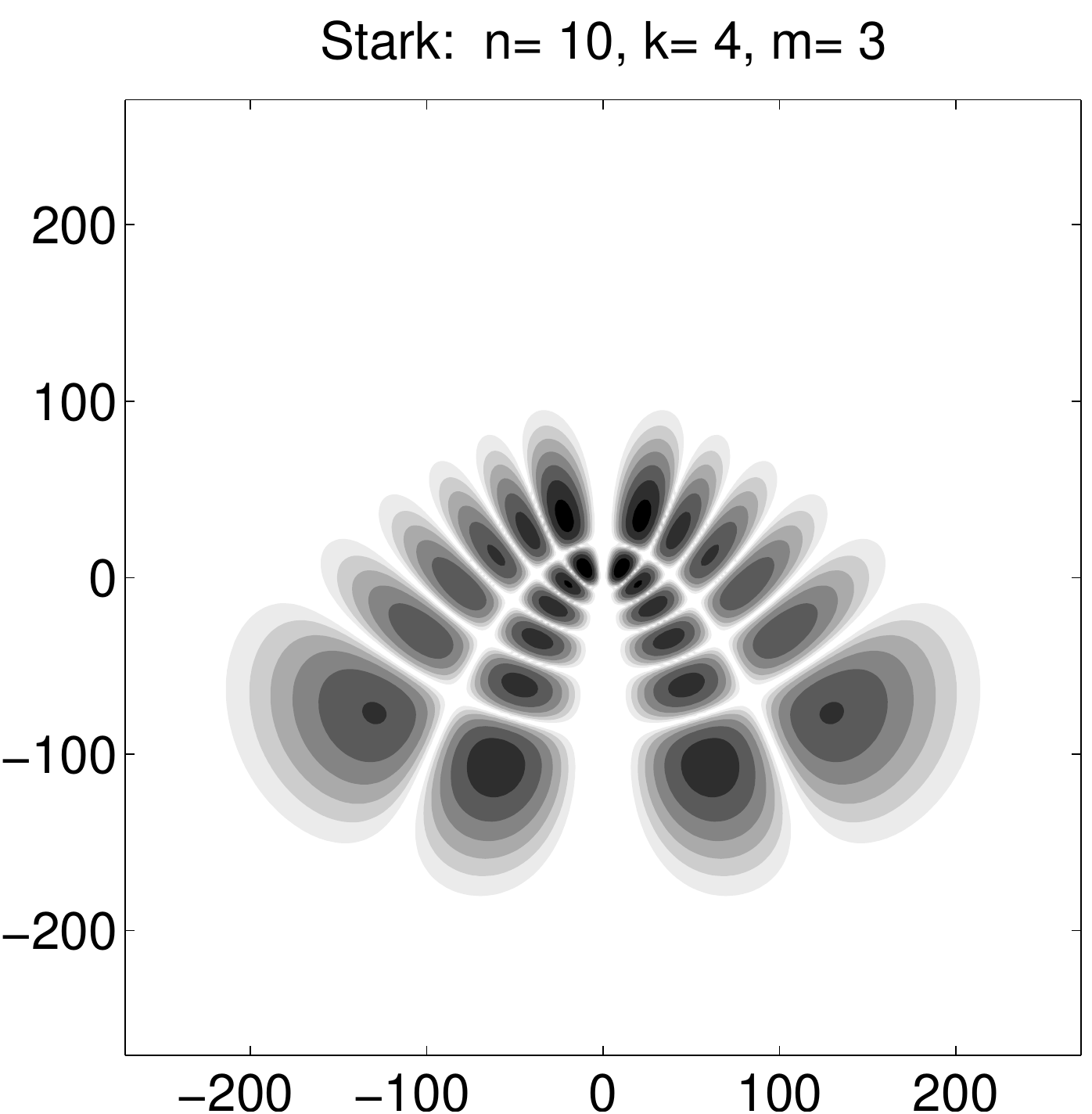} &
\includegraphics[width=\figdefdim] {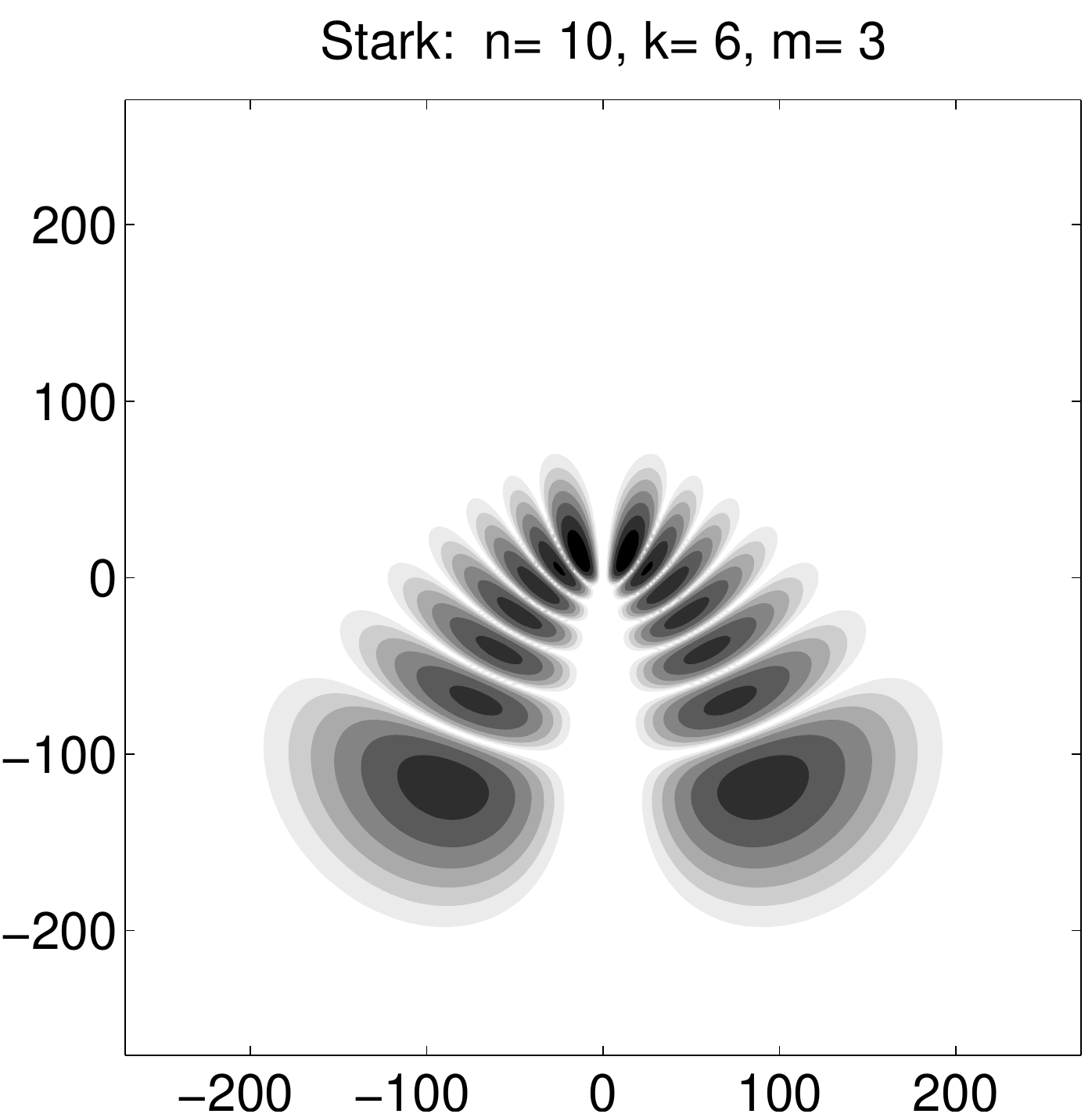} &
\includegraphics[width=\figdefdim] {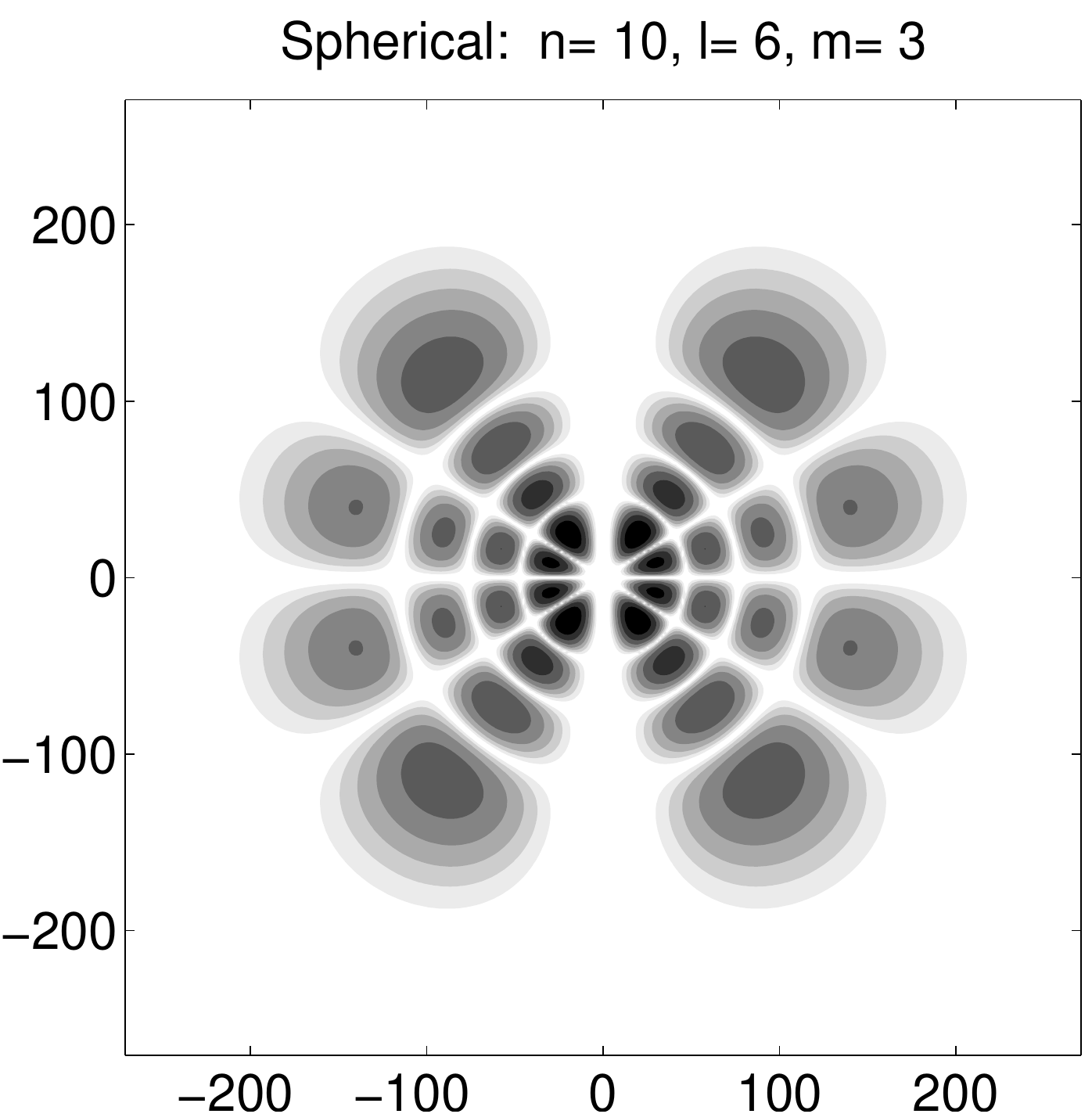} \\
\includegraphics[width=\figdefdim] {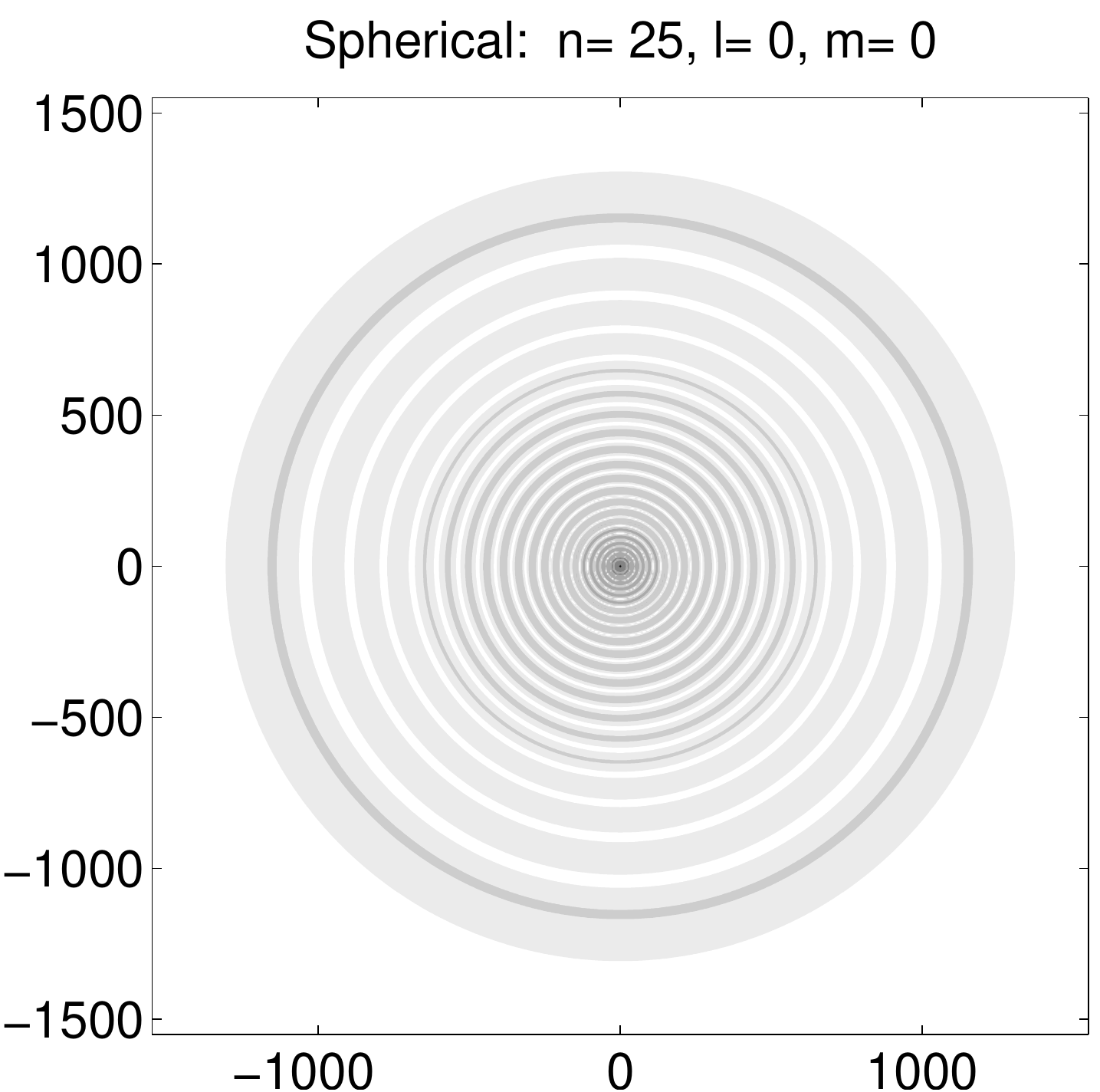} &
\includegraphics[width=\figdefdim] {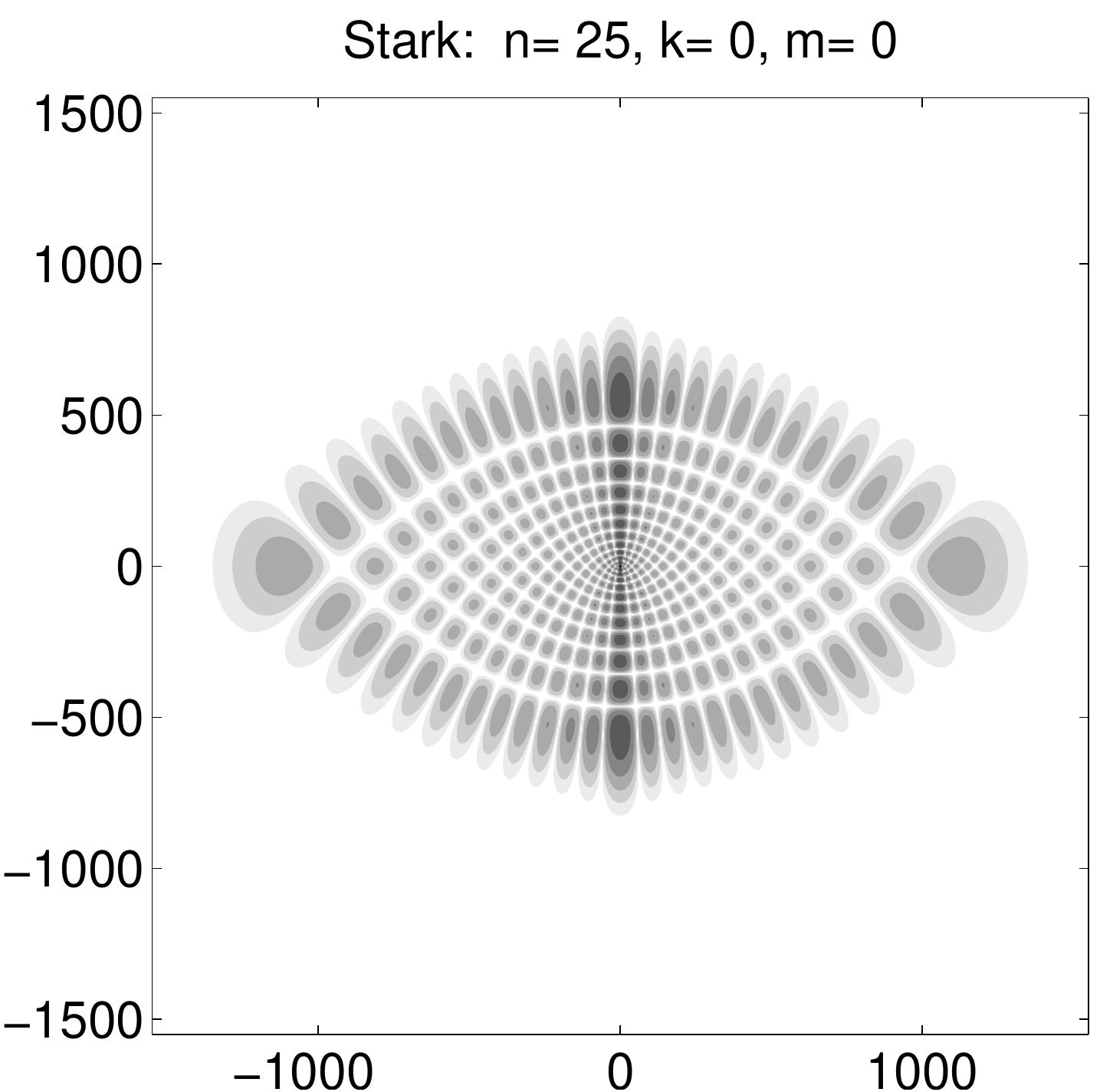} &
\includegraphics[width=\figdefdim] {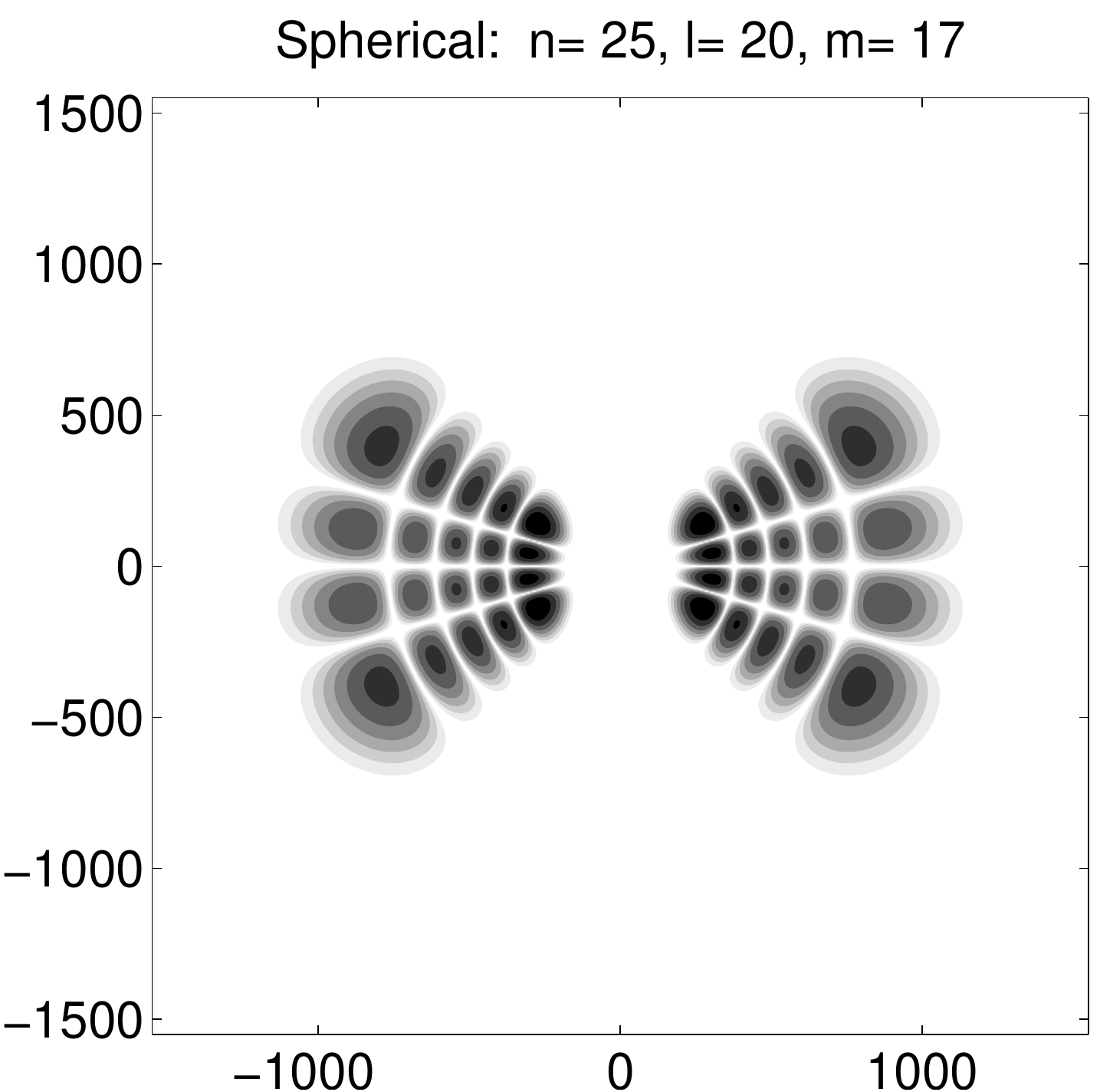} \\
\includegraphics[width=\figdefdim] {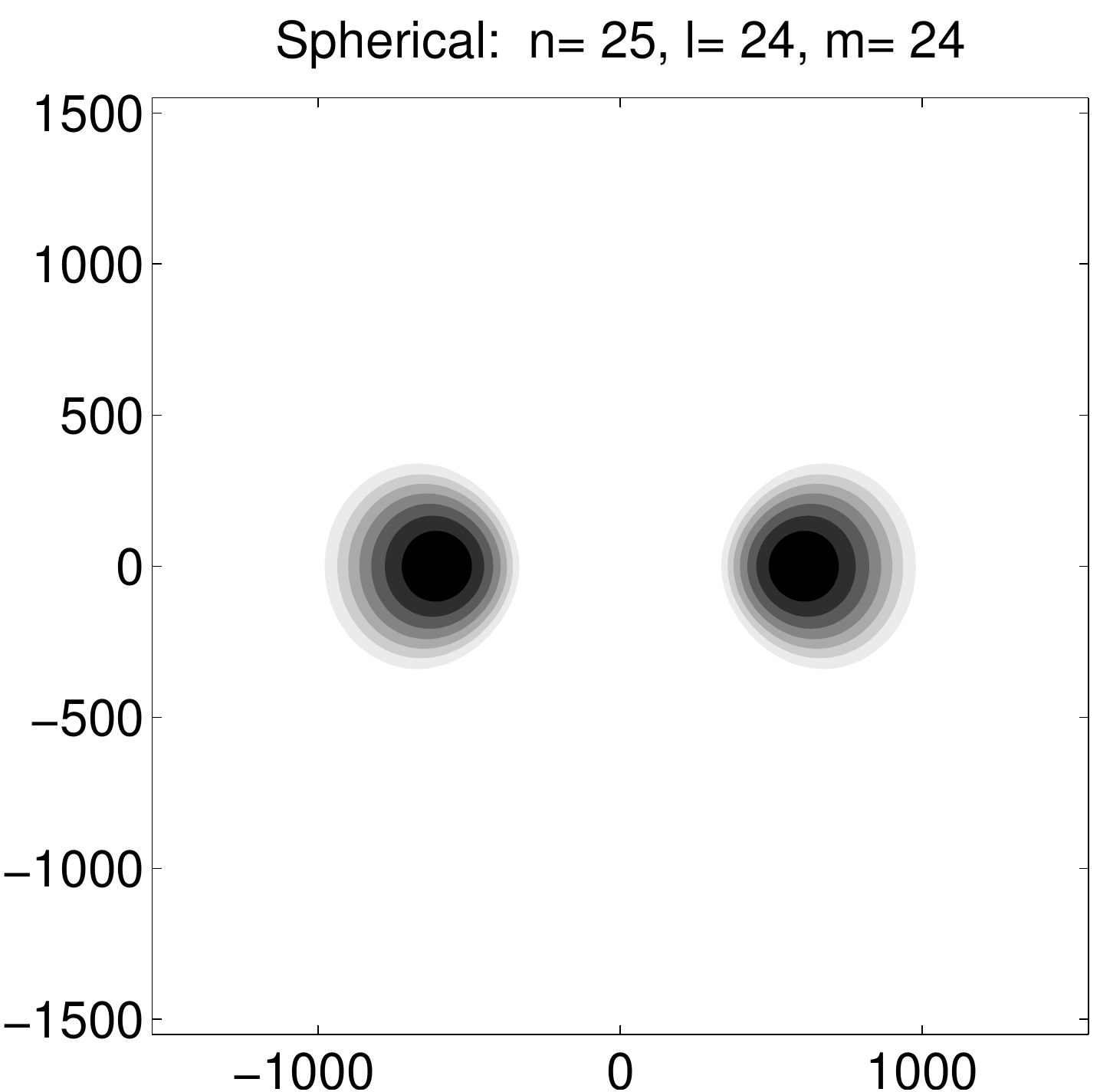} &
\includegraphics[width=\figdefdim] {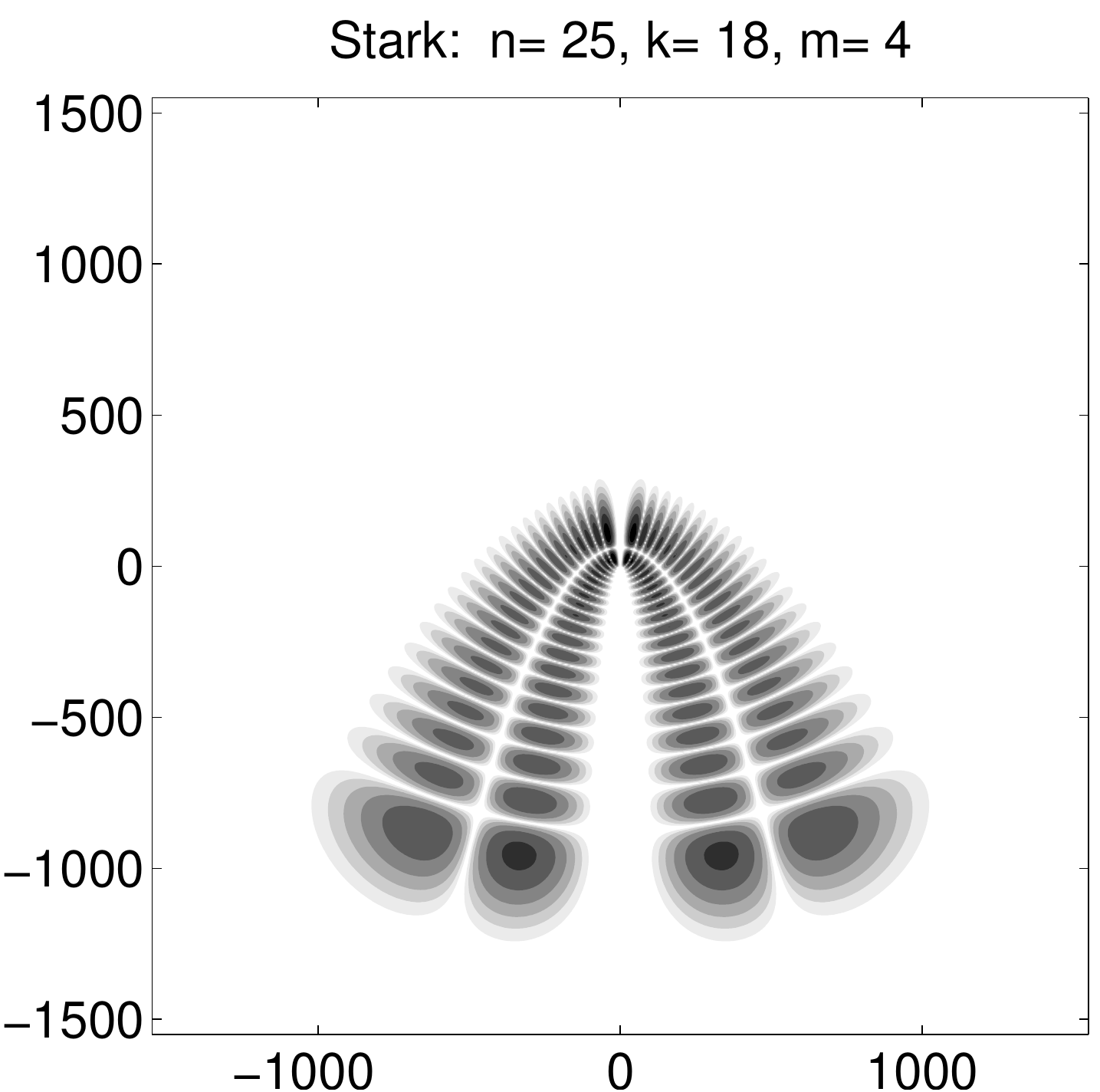} &
\includegraphics[width=\figdefdim] {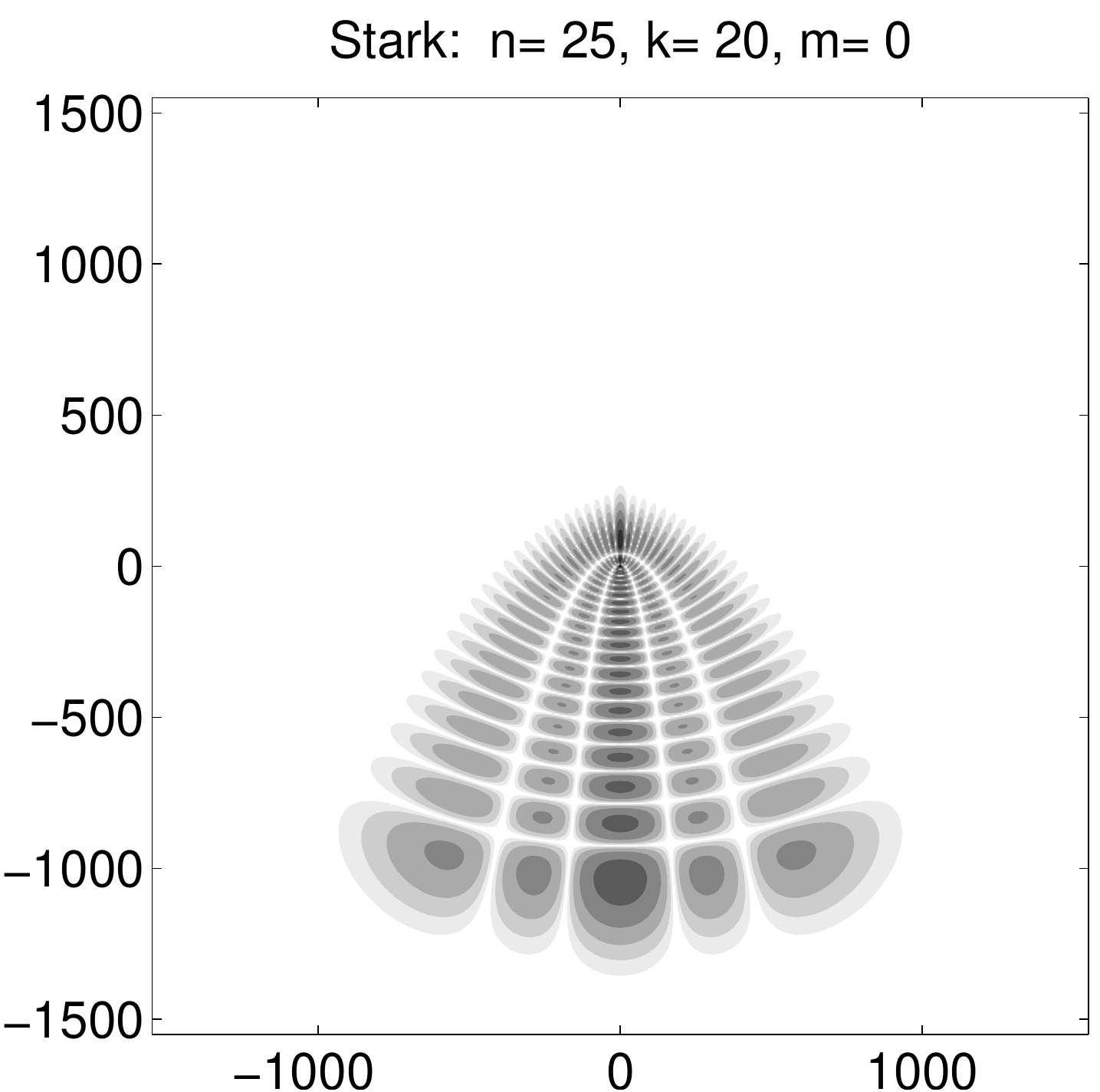} \\
\includegraphics[width=\figdefdim] {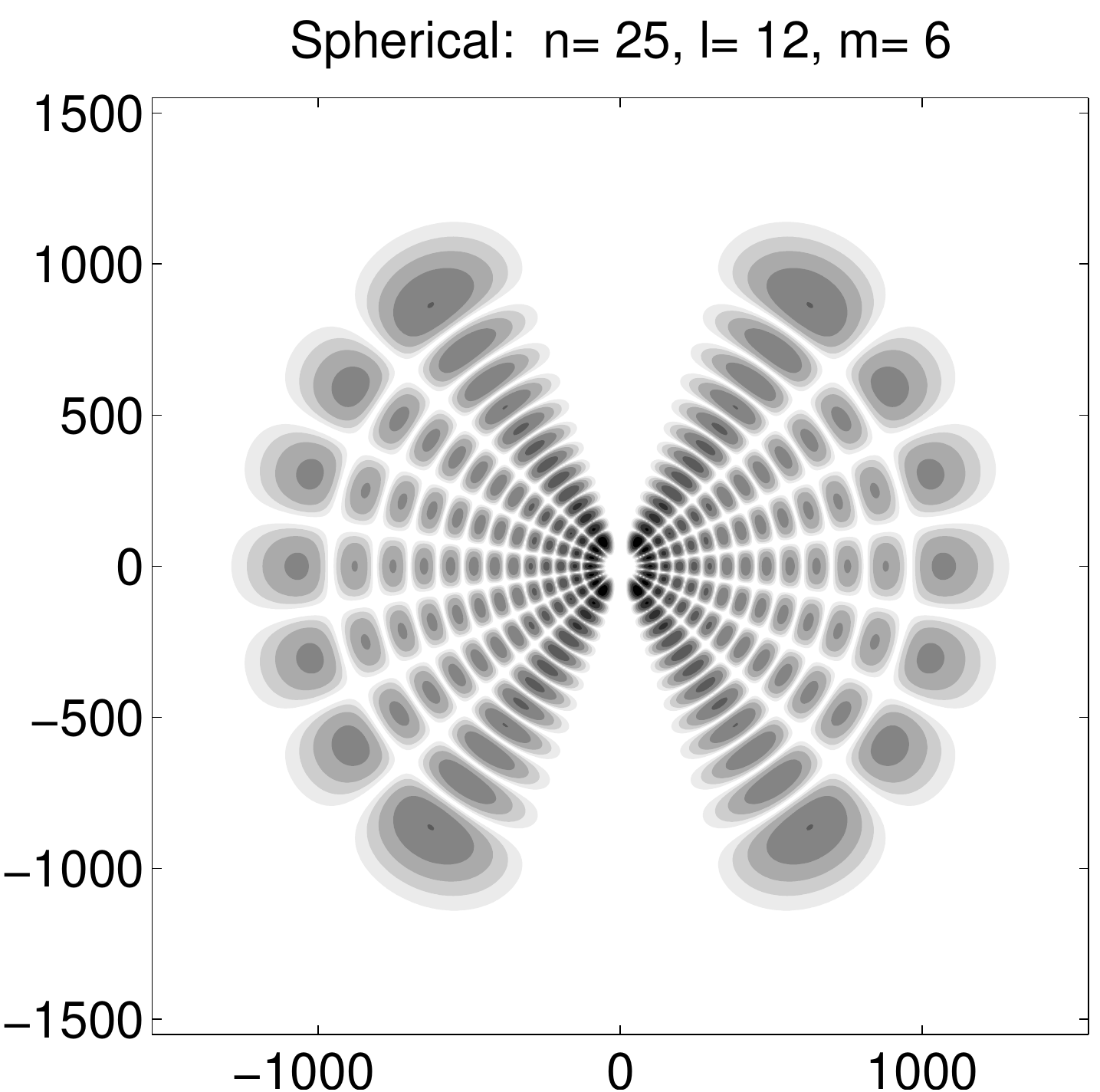} &
\includegraphics[width=\figdefdim] {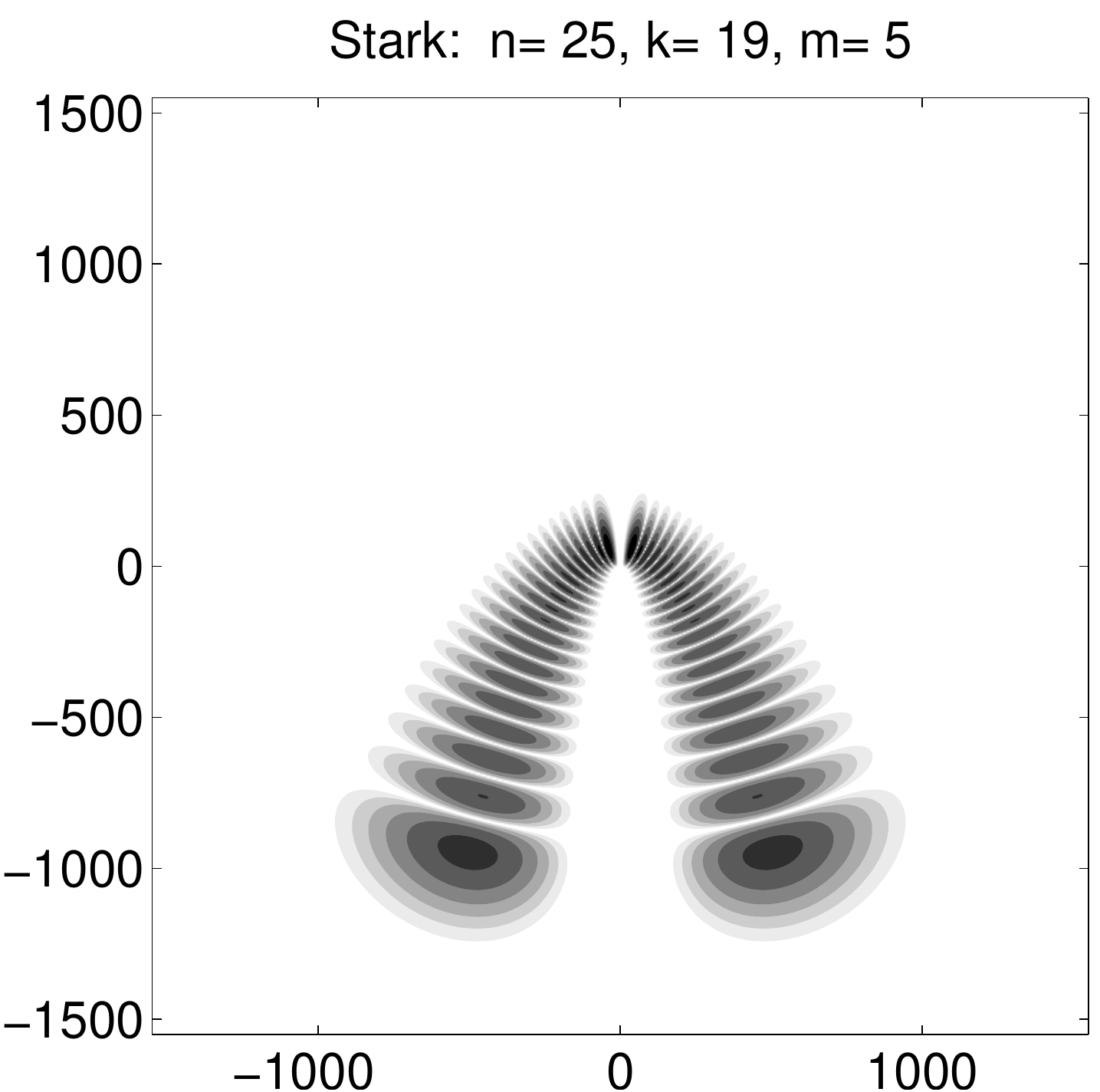}  &
\includegraphics[width=\figdefdim] {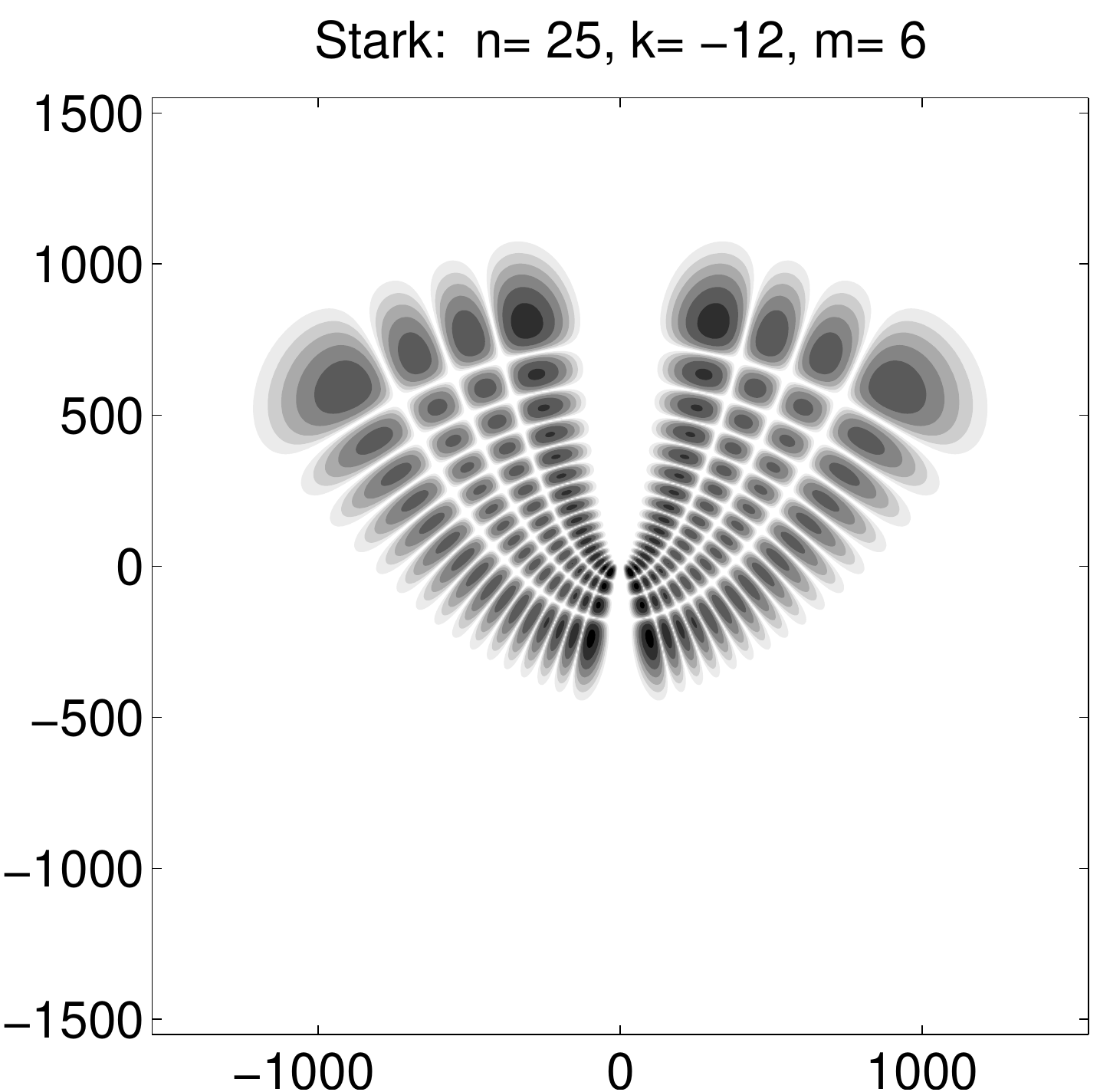}   \\
\end{tabular}
\caption{Example plots of Rydberg states. Note that all the states are cylindricaly symmetric,
i.e. the full shape is a rotation along the vertical z-axis. The states $m=0$
will have a solid block along the z-axis, all the $m>0$ states have an empty
tube - or a sort of funnel along the z-axis. Note the tiny black spot 
on the nucleus in the
$25s$ state. Note also that generally there are high density regions close to the nucleous in most cases.
} \label{flatfig_exammples}
\end{center}
\end{figure}

The $m=0$ states are not the only possible states. For a given $n$, there 
are exactly $n$ states with $m=0$ while there are $n^2$ various Stark states.
As an example, we compare the states of $n$=5 (table \ref{compare_nlm_nkm_5} ) 
and $n$=4 (table \ref{compare_nlm_nkm_4} )in the two quantization schemes,
the spherical system and the parabolic system. 

 We show both of the tables, since they demonstrate one interesting feature: only for odd values of $n$ 
we can have the state $k=0$ $m=0$ which has both cylindrical symmetry ($m=0$)  and reflection symmetry 
along the z-axis ($k=0$). For $n$ even such state does not exist. 
The $k=0$ $m=0$ states for large values of $n$ have a very striking and unexpected structure - they are also reflection symmetric, but they only exist for odd $n$.
%
%
%
%
%
%
%
%
%
%
%
%
                 \section{        Unfortunate representations of Stark states \label{stark_unfortunate}        }
%
%
%
%
%
%
The first example is the famous figure 6.1 of Gallaher book discussed above, 
where the density is multiplied 
apparently by a power of $r$. This even without mentioning this multiplication.
Our figure \ref{flatfig_gallagher} shows our opinion of how the figure should have been
presented. It is interesting to analyze why the original author of the figure
chose this extra multiplicative factor so strongly modifying the shapes
(as shown in our figure {landscape-stark}). It is very probable that the main reason is simply the wish to
eliminate the uncomfortable nuclear maximum. Note that the height of this maximum relative to 
the last radial maximum (for density) scales roughly with $n^4$, while its volume is of the order
of one atomic unit of volume independent of $n$. The multiplication 
then propagates and leads to confusions and omissions, originally caused by
inadequate visualization approach.

The next unfortunate choice we refer to has been made by Greene et al 
\cite{trilobite} 
%
In this case the authors correctly describe what the plot really is,
quoting their figure caption:
"A cylindrical coordinate surface plot of the electronic probability density,
  $2\pi \rho |\psi(\rho,z,0)|^2$ and $2\pi \rho |\psi(\rho,z,\pi)|^2$ .... ".
The plot is a part of a physically very important and motivating paper. However, the chosen representation
is rather unfortunate because the central region along the internuclear axis
which in fact is a maximum appears as being a node. It is a node due to the multiplication
by the cylindrical $\rho$. Also, the authors even write ``trilobite resembling density'' which
further  strengthens the impression of flatness, while the object has cylindrical symmetry.

This unfortunate choice of representation has been continued to a recent paper in Science 
by W. Li {\it et al } \cite{homonuclear_polar}
where the same method is used, but unfortunately in this case again omitting to specify this. 
The text inside of the
figure even explicitly indicates an erroneous $|\Psi({\mathrm r},{R}_0)|^2$, 
not including the $\rho$, which
however is indicated as an axis denotation. This is clearly only an omission, but again the choice of the
visualization method is unfortunate. In this case even more so, since the
main character of the state is $35s$.

Similar example of unfotunate choice of interpretation 
for the next type of states will be discussed in section 
\ref{interpretCES}.

%
%
%
\section{Coherent Elliptic States  and other Coherent superpositions \label{Elliptic_section}}
%
%
%
%
%

\subsection{Pauli's treatment of hydrogen atom in electric and magnetic fields \label{pauli_el_mag_math}}
%
In 1926 Wolfgang Pauli presented the problem of a hydrogen atom in external fields~\cite{Pauli}. 
Here he showed that (originally in the matrix formulation of quantum mechanics) how the stationary states 
of the hydrogen atom can be treated in classified using a combination of 
angular momentum operator and the operator known today as Laplace-Runge-Lenz operator.
%
The Hamiltonian for atomic hydrogen 
in a combination of  electric ($ \nvec{E} $) and magnetic ($ \nvec{B} $)  fields is given by
\begin{equation}
      H = H_0 + \nvec{E}\cdot \nvec{r} + \frac{1}{2} \nvec{B}\cdot\nvec{L}, \label{Eq:Hamiltonian_semiclassical}
\end{equation}
with
\begin{equation}
 H_0 = -\frac{ \nabla^2}{2} - \frac{1}{r}, \label{Eq:hamiltonian_basis}
\end{equation}
being the Hamiltonian of an unperturbed hydrogenic atom, and the angular \mbox{momentum} operators of the electron is 
given by $\bf L = \bf r\times \bf p$.\\
When the dynamics is restricted to a single hydrogenic $n$-shell the Pauli's operator replacement can be 
%
\begin{equation}
   \nvec{ r}=\frac{3}{2}n  \nvec{  A   },
\label{Eq:H3}
\end{equation}
where \textbf{A} is the quantum mechanical counterpart of the classical Runge-Lenz vector,
\begin{equation}
 \nvec{A}=\frac{1}{\sqrt{-2H_0}}
   \left[ \frac{1}{2}(\nvec{p}\times\nvec{L}-\nvec{L}\times\nvec{p})-\frac{n}{r} \nvec{r} \right].
   \label{Eq:Runge-Lenz-vector}
\end{equation}
Pauli recognized that one can combine the operators $\nvec{L}$ and $\nvec{A}$
\begin{eqnarray}
\nvec{J}_1 &  =  &   \frac{1}{2}(\nvec{L} + \nvec{A})~,\nonumber \\
\nvec{J}_2 &  =  &   \frac{1}{2}(\nvec{L} - \nvec{A})~,
  \end{eqnarray}
to two basically independent operators which obey the commutation
relations of angular momentum, and can be used as independent operators.
However, already Pauli have shown that these two operators are not
fully independent. For states inside one $n$-shell their squares must have the
same eigenvalue, $J_i^2 \rightarrow  j_i(j_i+1), j_i=\frac{n-1}{2}$.

%
Then, the Hamiltonian of the intrashell system can be transformed
using the pseudospins operators and the energy 
replacement, $H_0=-1/2n^2$ it gets the form
\begin{equation}
	H= H_0 + {\bm \Upomega}_1\cdot   \nvec{ J}_1 + {\bm \Upomega}_2  \cdot  \nvec{ J}_2,
\label{Eq:H6}
\end{equation}
The vectors ${\bm\Upomega}_1$  and ${\bm\Upomega}_1$ are also given by a simple
relation
\begin{eqnarray}
{\bm\Upomega}_1 &  =  &   \frac{1}{2} \nvec{B} + \frac{3}{2}n \nvec{E},\nonumber \\
{\bm\Upomega}_2 &  =  &   \frac{1}{2} \nvec{B} - \frac{3}{2}n \nvec{E},
\label{quasispins-proj}
  \end{eqnarray}
and resemble the role of two  effective magnetic fields, since they appear in the same relation as $\nvec{B}$ and 
$\nvec{L}$
in equation (\ref{Eq:Hamiltonian_semiclassical}).

In equation  \ref{quasispins-proj} the two operators $\nvec{ J}_1 $ and $\nvec{ J}_1 $ 
are defined as two vector operators, and from the definitions 
it follows that their components can be chosen with respect to any set of 
axes (\ref{Eq:H6}). 

Diagonalization of the transformed hamiltonian eq. (\ref{Eq:H6}) is thus elementary. 
If we chose as basis states which at the same time are eigenstates of 
$J_{1z'}$ and $J_{2z''}$ where axis $z'$ is in the direction of ${\bm\Upomega}_1  $ 
and ${z''}$ the direction of ${\bm\Upomega}_2  $ . We do not know these states
$  | m_1 m_2   \rangle$  
explicitely, but they are defined by the property 
$$
  J_{1z'}  | m_1 m_2 \rangle = m_1  | m_1 m_2  \rangle 
  \ \ \ \ \ \ \ \ \ \ \ \ \ \ \ \ \ \ \
   J_{2z''}  | m_1 m_2 \rangle = m_2  | m_1 m_2  \rangle 
$$
Generally, for a given E and B these quantum numbers are not
related in any way to the quantum numbers $m$ or $k$ discussed in section \ref{stark_section}.
\begin{equation}
   E_{m_1, m_2}=m_1  \Omega_1  +  m_2 \Omega_2
\end{equation}
where $ \Omega_1$ and $\Omega_2$ are the "lengths" of the two vectors in 
equation \ref{quasispins-proj}. When B and E are perpendicular, these two constants are equal
and we obtain a set of states which have a structure similar 
to table \ref{Pauli_nlm_nkm_5} single state on the top and botom with equidistant eigenvalues
with increasing degeneracy by one from bottom towards the middle and again decreasing
towards the top. The $n^2$ states plotted
at positions of their eigenvalues form a diamond with middle width $n$ and height $n$.

When E and B are not perpendicular the above equations are still valid but the
eigenvalues do not
show any particular symmetry and the degeneracy is mostly lifted.

\subsection{Interpretations of CES }

The two extreme states $|m_1=-\frac{n-1}{2}, m_2=-\frac{n-1}{2} >$ and $|m_1=\frac{n-1}{2}, m_2=\frac{n-1}{2} >$
have been called coherent states, from the point of view that they have minimum
uncertainties of the involved operators. 
These coherent elliptic states (see e.g. Bommier et al \cite{Bommier_book} are 3-dimensional objects, in the sense that
their densities varies in all three dimensions with only one symmetry retained - the reflection symmetry.
In order to visualize  such states one then needs to represent
a function of three variables in the two-dimensional geometry of the paper or screen - 
it means some  three-dimensional help-objects must be used. There is a well established method
of isosurfaces, i.e.  plotting surfaces where the functional value remains the same. 
This is an extension of the contour plot used above to 3 dimensions. 
Often only one value is used, in our case a small value whose isosurface would enclose 
all the higher values of density.

This method we shall be using in the following, with some variations.
Very useful method is to use several isosurfaces, using both transparency of the objects and cuts
or slices.
Technically, these methods are available in many packages, as already mentioned
we use the package MATLAB. General feature of all such packages is that they
should not be used with the standard parameters, adjustment of parameters is an essential
part of the visualization work. 

Plots recognizing this and somewhat resembling our type of plots (with 1990's graphics) have been presented
for a somewhat similar problem of 
elliptic coherent states for two dimensional harmonic oscillator in  ~\cite{Pollet}. 
The term coherent is also used in the same sense, 
when the quantum fluctuations are as small as possible considering the uncertainty relations 
for observables related to the orientation and the eccentricity $e$ of an ellipse. 
These states are simple to construct and their wavefunction is localized on the corresponding 
classical elliptical trajectory. 
With increasing principal quantum $n$, the localization on the classical 
invariant structures is more pronounced.

\begin{figure}[h]
\begin{center}
\begin{tabular}{ll}
\includegraphics[width=6cm] {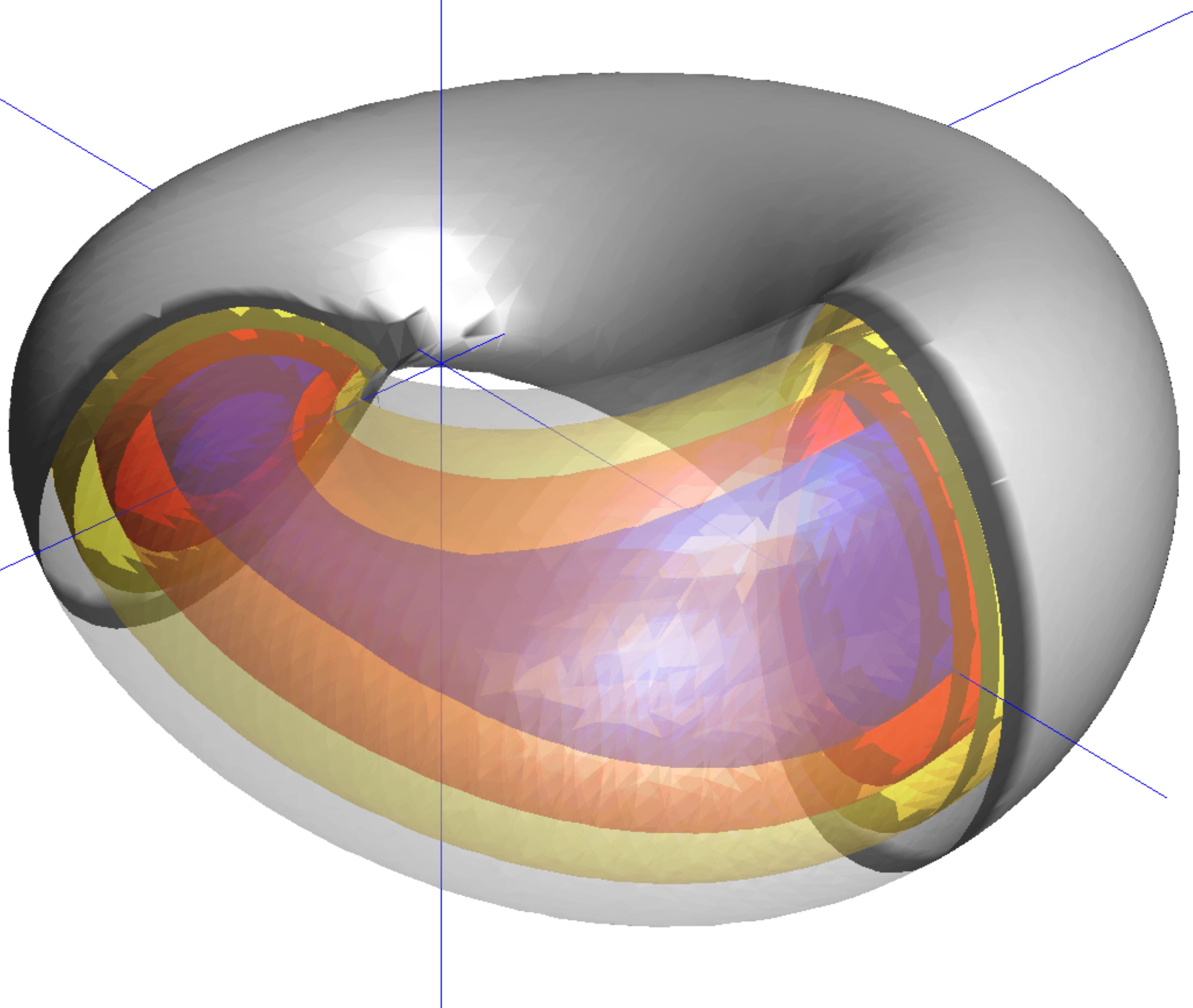} &
    \includegraphics[width=4.8cm] {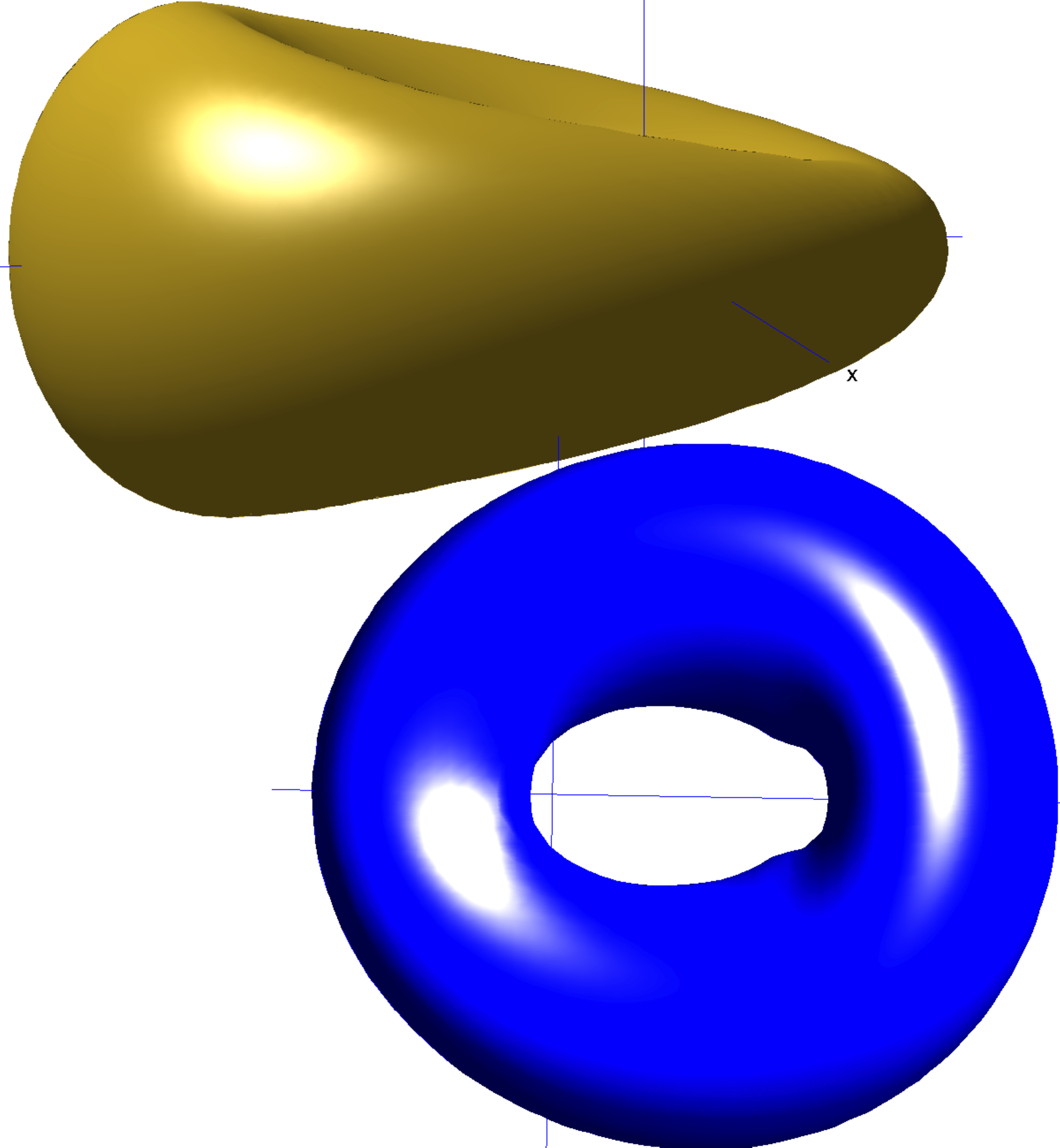} \\
\end{tabular}
 \caption{Elliptic state with $n$=11 and Bommier's $\alpha=0.583$. Electric field is not too strong.
The 
evaluated eccentricity is e=0.55.
\label{ellipse_alpha_0_583} } 
\end{center}
\end{figure}

\begin{figure}[h]
\begin{center}
\begin{tabular}{ll}
\includegraphics[width=6cm] {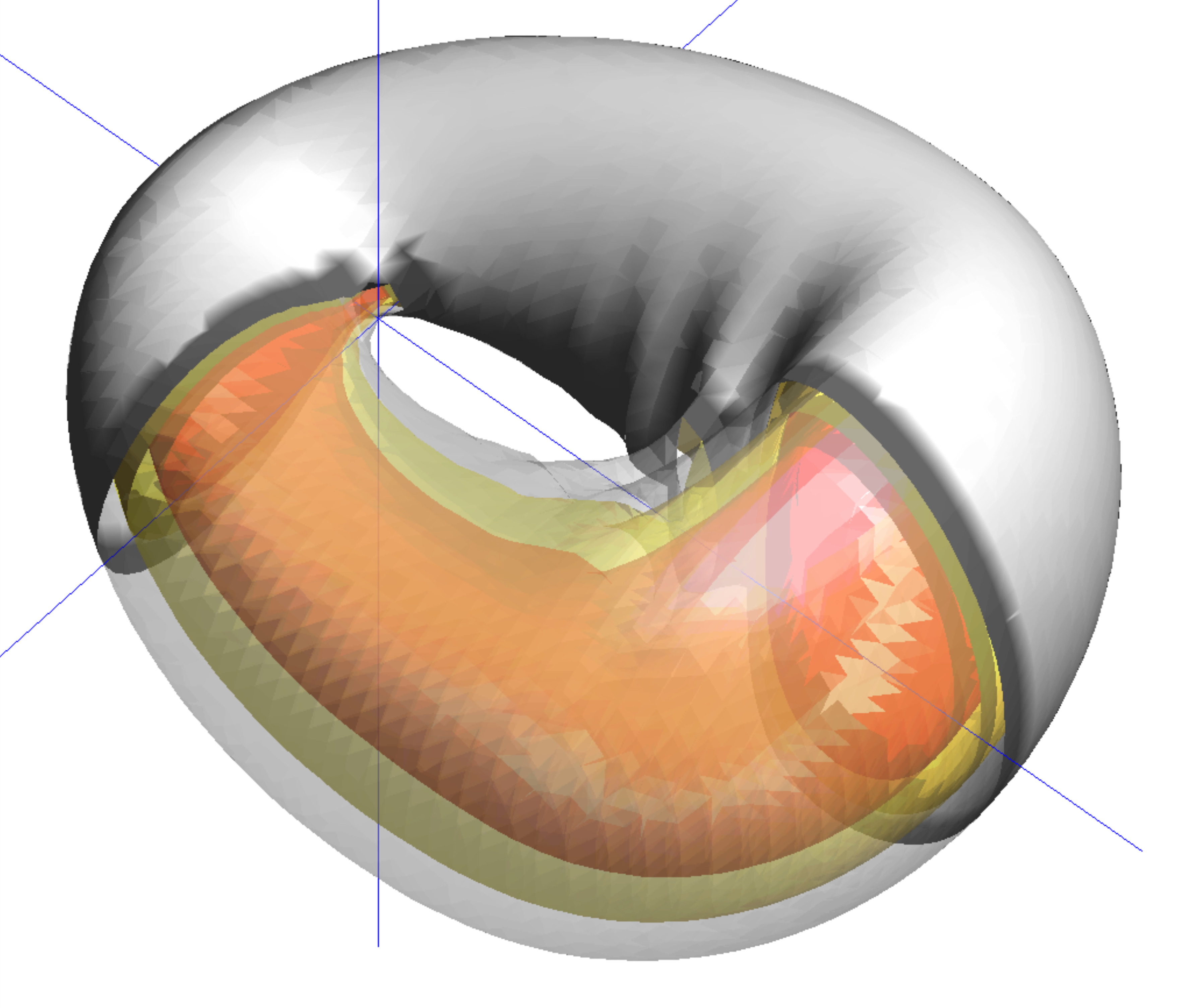} &
    \includegraphics[width=4.8cm] {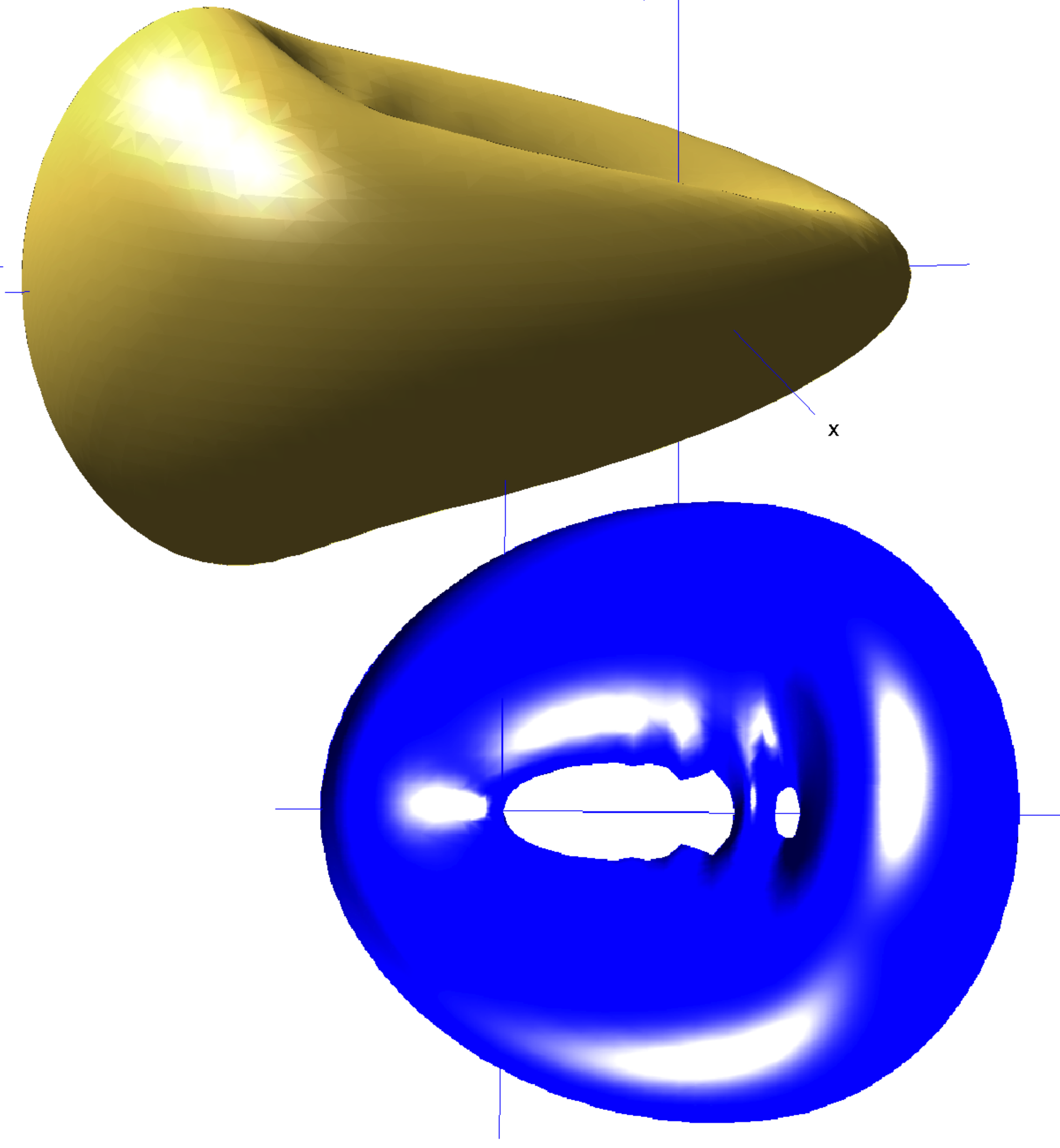} \\
\end{tabular}
\caption{Elliptic state with $n$=11 and Bommier's $\alpha=0.8$. Electric field is quite strong 
and the ellipse starts to get features of the Stark states (the extra rib). The 
evaluated eccentricity is e=0.71.
\label{ellipse_alpha_0_8} } 
\end{center}
\end{figure}

We start here by presenting our version of representation of Bommier's et al CES 
with the help of isosurfaces.
Figure \ref{ellipse_alpha_0_8} shows a relatively strong electric field, but in fact
one only leading to eccentricity, $e=0.71$. Here the ellipse starts to get features of the Stark states  
apparent here as the extra rib. Figure \ref{ellipse_alpha_0_583} shows a typical 
CES for eccentricity  $e=0.55$.

%
%
%
%

Probability density current is present in the circular states, i.e. in the states which are 
made possible by magnetic field. If we think about a possible preparation of such state,
exposure to magnetic field defines its direction as a quantization axis and orders the
states according to their magnetic quantum numbers, i.e. eigenstates of ${\bf L} \cdot {\bf B}$.
The two circular states with maximum possible absolute value of($n-1$) B  are also 
eigenstates of $L^2$, while all the other states are not eigenstates of $L^2$. They are degenerate
and can be chosen to be classified by $L^2$ as in the table \ref{compare_nlm_nkm_5}. 
Similarly, the Stark states can be chosen by the electric field in the z-direction, and then 
they can be additionally characterized by the $m_z$ quantum number, as shown in table \ref{compare_nlm_nkm_5}.

However, only when the electric and magnetic actions are perpendicular one can obtain
states which approach the elliptic states.
Construction of all the $n^2$ states is straightforward by numerical diagonalization.
Bommier et al have analyzed in great detail the analytic construction of 
the lowest state, which is coherent in the sense that it minimizes the variation 
in $L_z$ and $A_x$. This aspect has been discussed already by Pauli, but not
to the same depth. In perpendicular crossed fields, when B definez the z
direction, the two extreme states change between
circular (plus and minus orientation) in the x-y plane to linear Stark in positive and
negative x-direction (E defines x-axis).
Bommier et al derived an analytic expression for these states in terms of
a summation over the $(n,l,m)$ states.

In discussion of these states a serious misunderstanding has been propagated from
the earlier discussed cases, i.e. that the densities must be transformed in
order to avoid intuitively uncomfortable features.

%
%

\section {Interpretations of Coherent Elliptic States \label{interpretCES}}

The coherent elliptic states have been discussed by Bommier et al,  in several papers.
%
%
\begin{figure}[h] 
\begin{center}
\begin{tabular}{ c c }
 \includegraphics[width=7cm]{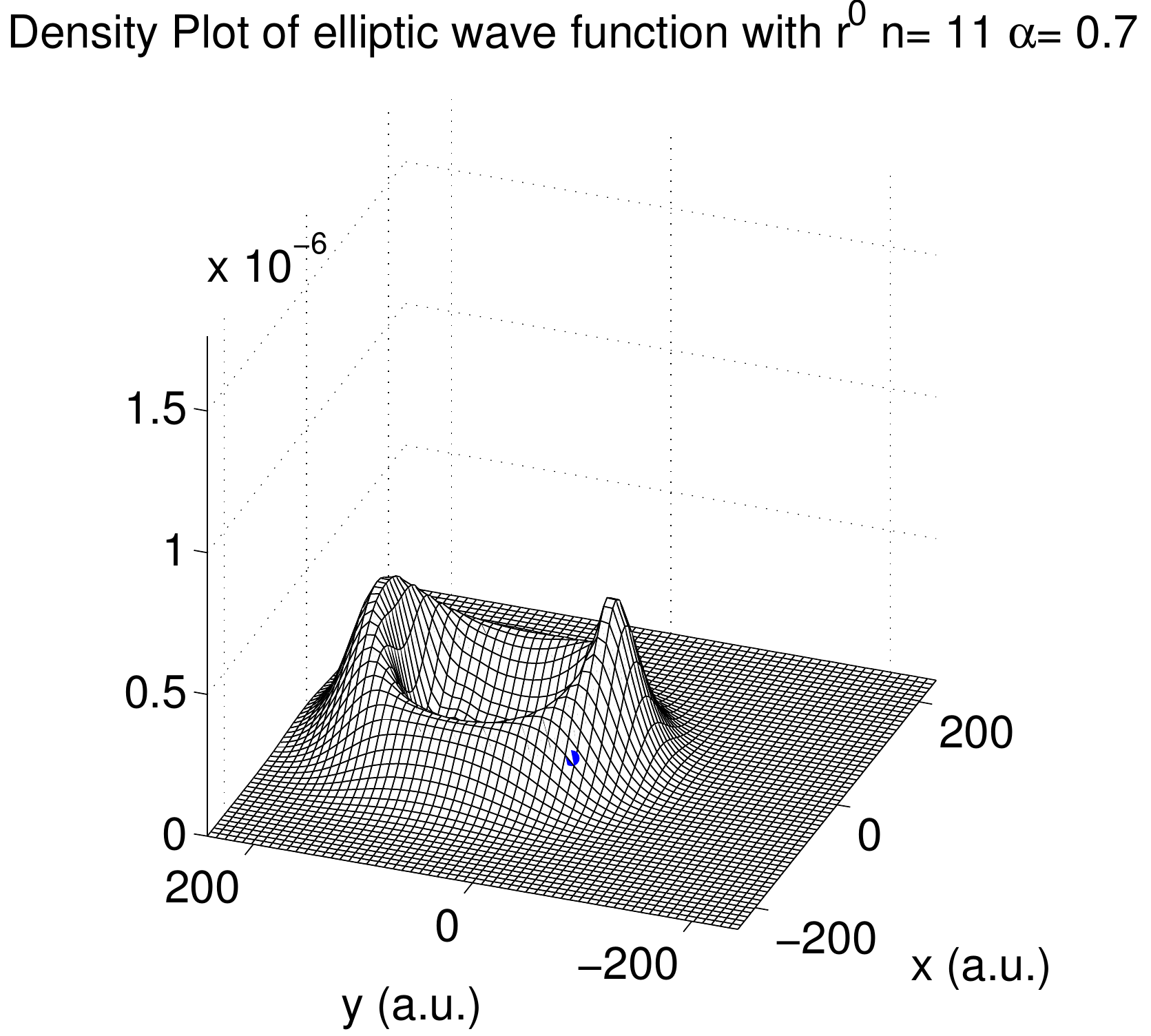}  
   &  
 \includegraphics[width=7cm]{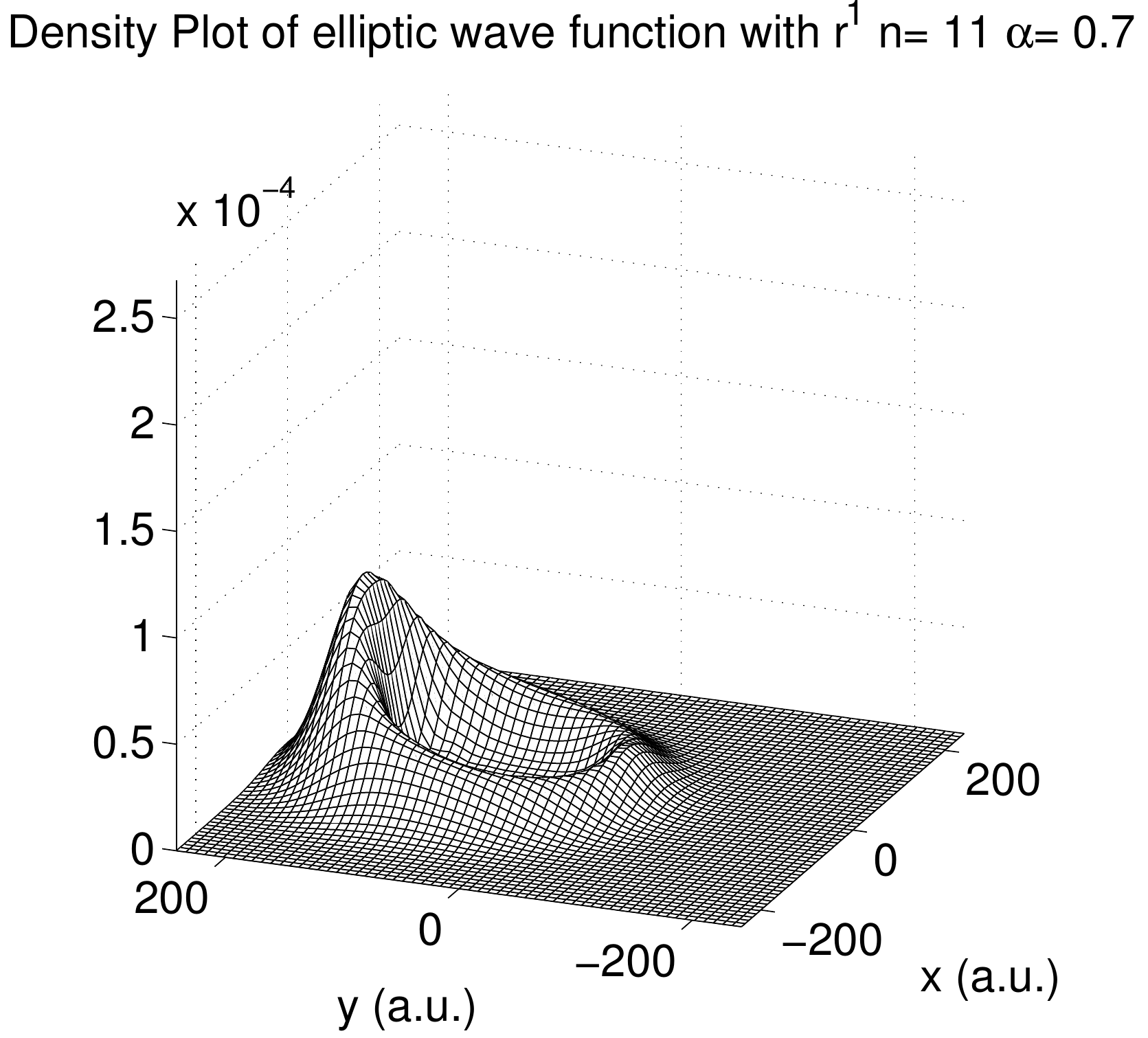}  
 \end{tabular}
\caption{Elliptic state with $n$=11 and Bommier's $\alpha=0.7$. The actual density
shown to the left has maximum both at aphelion and at perihelion.
The right picture shows one possible transformation
attempting to avoid the peak at perihelion.\label{twodim_ellipse} } 
\end{center}
\end{figure}  
%
%
%
In 1989 J.-C. Gay, D. Delande  and A. Bommier 
\cite{Bommier}
introduce the {\it Atomic quantum states with maximum localization on classical elliptical orbits}
On their figure 1(a) they show a plot of the electronic density in the (x, y) plane, 	
showing	localization on the ellipse. 
In Figure 1(b) they show a very  a very illustrative  cut in the perpendicular (z, x) plane showing that 
the elliptic state is strongly localized near the z =0 plane, a very nice feature
to some degree showing the wedge shape of the elliptic states.
%
%
%
\begin{figure}[h]
\begin{center}
\includegraphics[width=8cm]{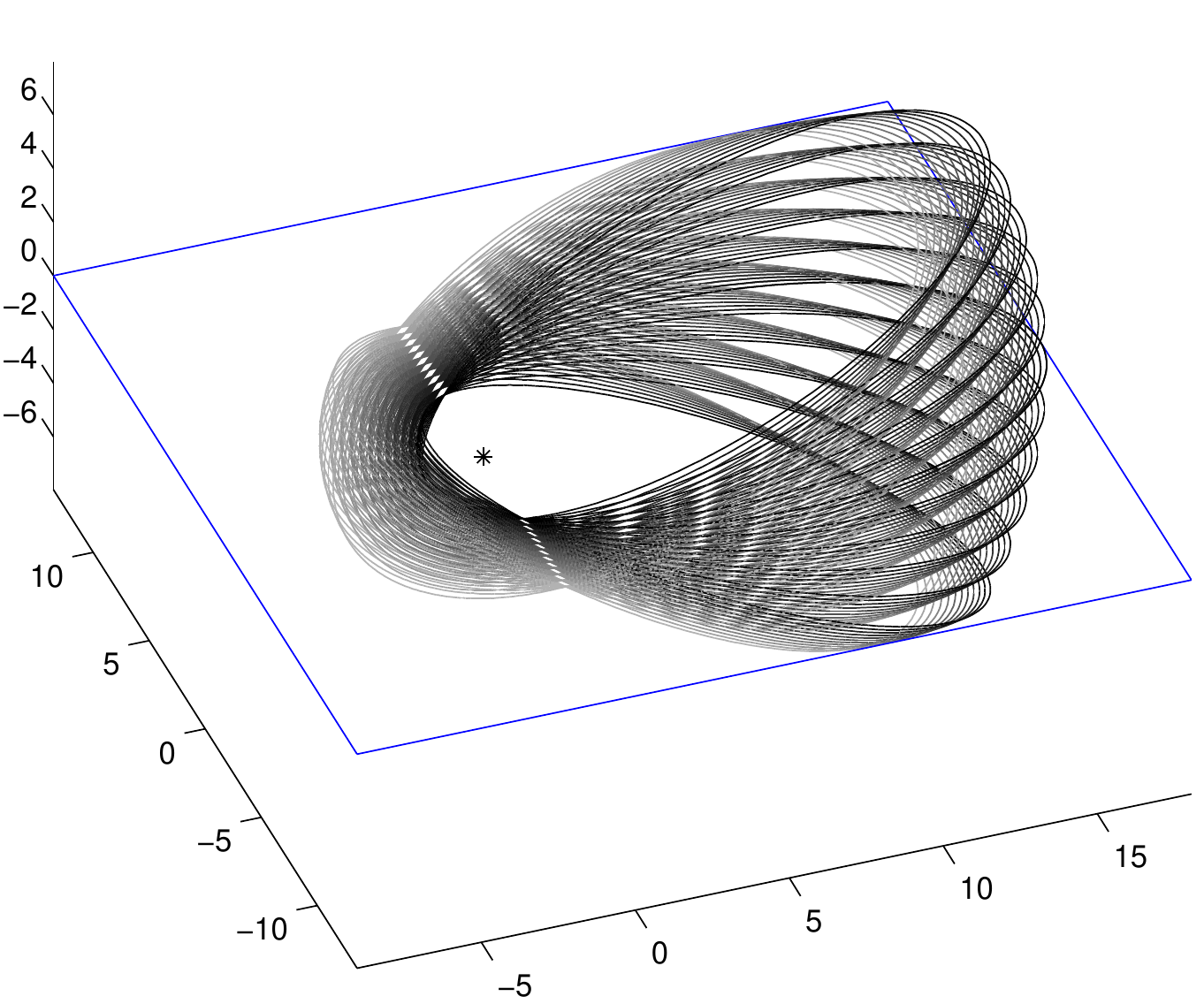} \\
\caption{Classical elliptic trajectories with angular momentum uncertainty. This illustration 
displays increased density at the perihelion. 
\label{ellipse_classical} } 
\end{center}
\end{figure}

Then they add a completely unnecessary figure 1(c), (equivalent to 
the right part of our figure  \ref{twodim_ellipse}
which they describe as ``{\it ... represents the electronic density in the (x, y) plane 
after averaging over z motion.
 It is still localized on the ellipse and the distribution is peaked at aphelion 
(minimum velocity) as expected from semiclassical arguments.	
The peaking	at the perihelion	
(maximum velocity) which Fig. 1(a) exhibits is smoothed out}''.

The smoothing is once more mentioned also in the figure caption,
there as  ``{\it The maximum at perihelion has been smoothed out. 
The one at aphelion is expected from semiclassical arguments. }''

There is absolutely no need to smooth out the maximum at the perihelion. 
This maximum is present in all the elliptic trajectories, even 
for the states discussed in section~\ref{Crossed_fields_section}.
The authors did not like this maximum and tried to get rid of
it by modifying the results.

Many scientists have since then been influenced by  this 
misunderstood necessity of smoothing out the unwanted maximum.
We refer here to one more example of an important review
co-authored by one of our close collaborators,
the review of {\it Charge transfer from coherent elliptic states}
\cite{Charge_transfer}. In their figure 4  the authors show
the "smoothed" density introducing it as
{\it The probability 
distribution projected onto the plane of the orbital is 
concentrated near a classical elliptic path.}, i.e. now the {\it smoothing}
is described as a {\it projection}. 
Further the caption of the figure states that
{\it The electron is more likely to be found at aphelion far 
from the nucleus, at the left side along E , 
than at perihelion close to the nucleus, at the right side.}

The last statement is true, but can not really be seen from the figure. 
It seems that it is in fact an attempt to formulate in physically correct terms
the original smoothing argument for removing the perihelion density maximum.

The figure has no other role in this important review paper than an illustration,
but it is a pity that this illustration repeatedly propagates
an unfortunate misunderstanding.
In our figure \ref{ellipse_classical} we sketch one possible 
illustration of the perihelion maximum. All the classical elliptic trajectories
corresponding to elliptic paths displaying the uncertainty in the 
angular momentum - and thus present
in other (tilted )planes than the x-y plane, 
must pass through a rather narrow region close to
the perihelion of the ellipse in the x-y plane. This can lead to a maximum,
even if the higher velocity for each individual orbit would lead to a minimum.
\begin{figure}[h] 
\begin{center}
\begin{tabular}{ c c }
\includegraphics[width=7cm]{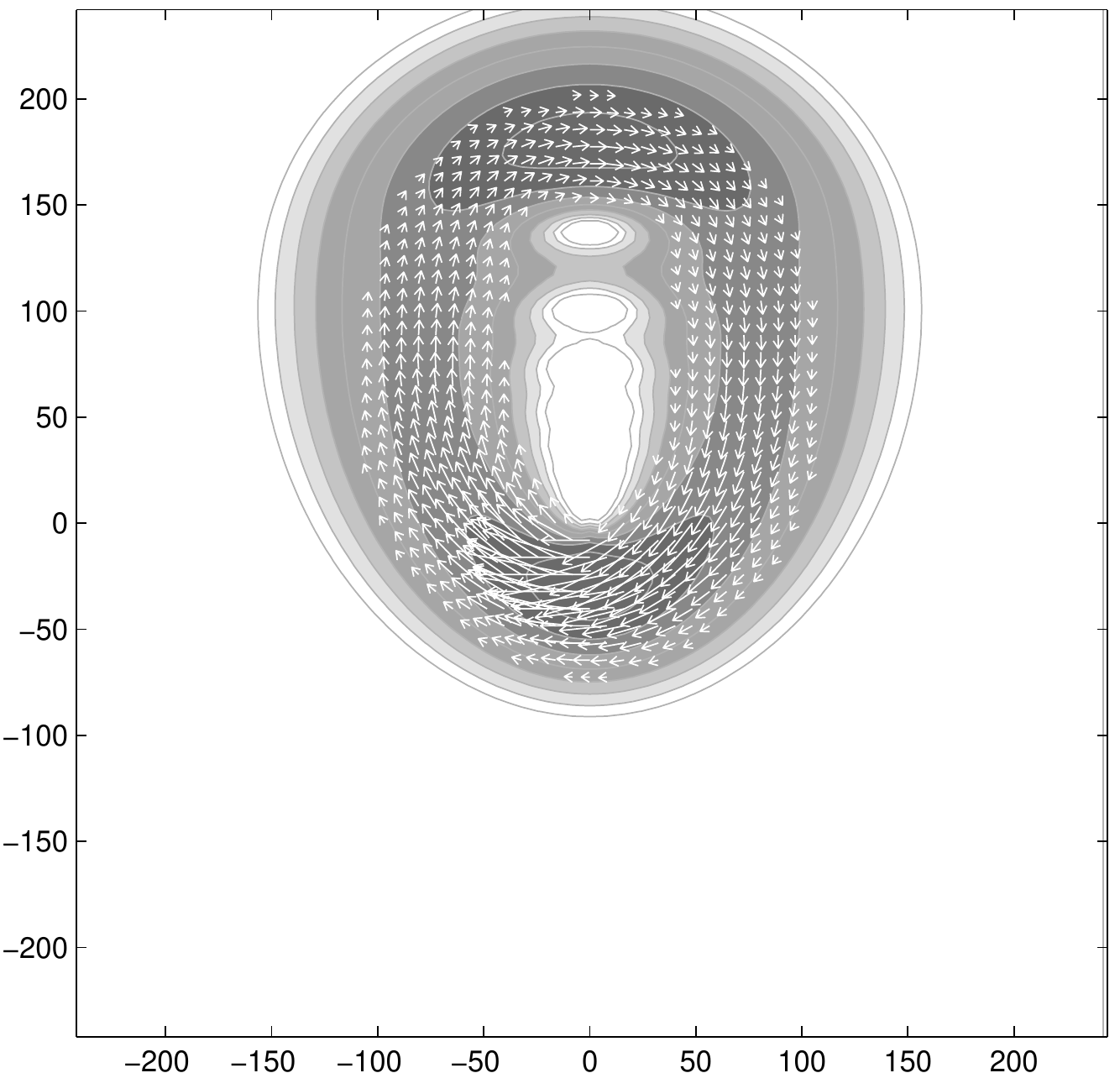}
&
 \includegraphics[width=7cm]{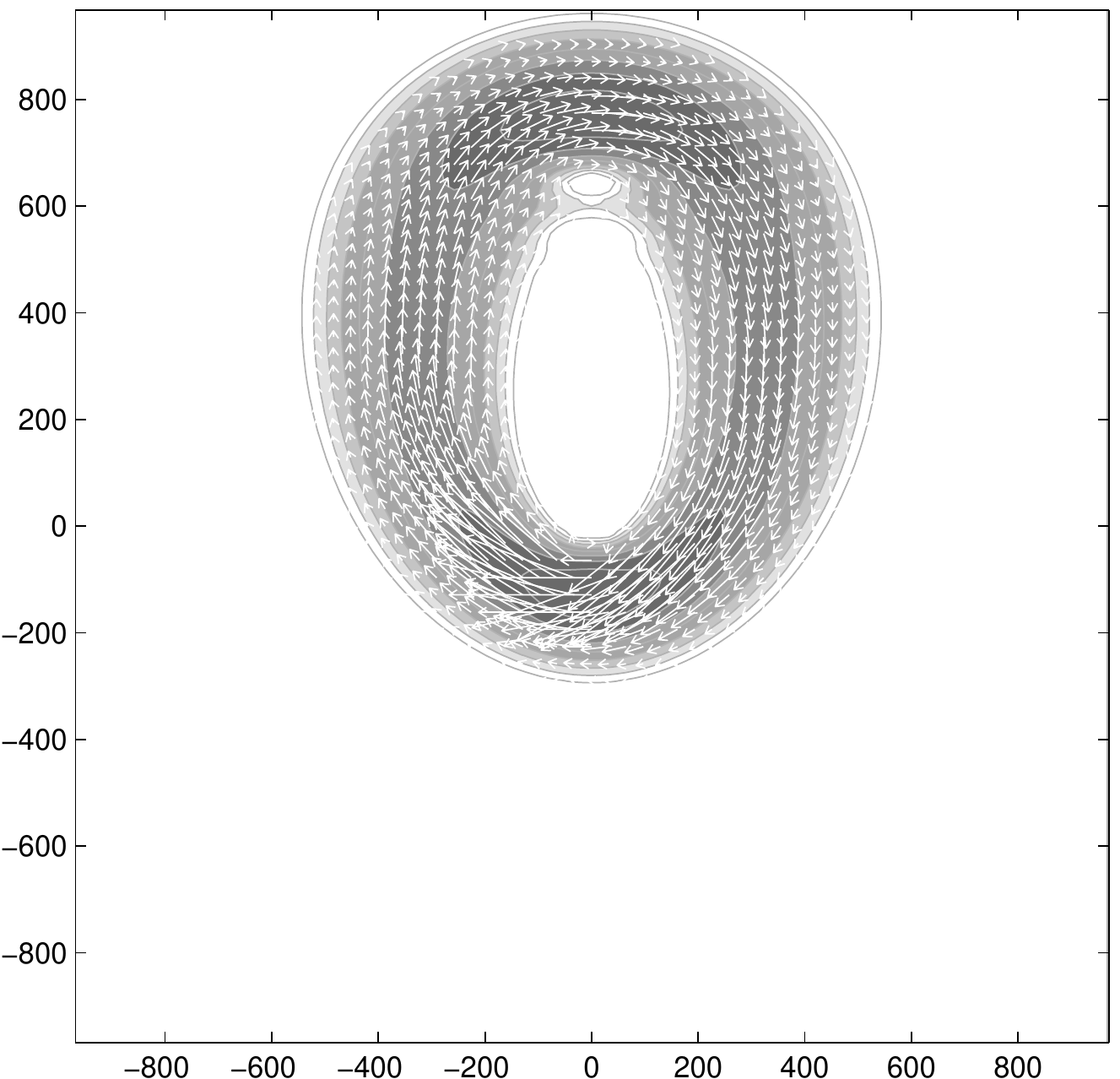}   \\ 
\end{tabular}
\caption{Elliptic states for $n=11$ and $n=22$. Plotted is probability density
in the x-y plane as for the plots in figure. The white arrows 
show the probability density current. The arrows are shown only
for values larger than interactively set threshold. The length
of the arrows is in a nonlinear way proportional to the
current density. The plotting method has been developed in
our studies \cite{comp_graphics} of atomic collisions in the 1990's.
\label{ellipse_current} } 
\end{center}
\end{figure}

This argument, however might not be too precise. To provide indication
of what else could be done with the interpretation of these
coherent states we have added simple diagrams of  
probability density $\rho$ and probability density current $\nvec j $
\begin{equation}
\rho(\nvec{r},t)=\psi^{*}(\nvec{r},t)\psi(\nvec{r},t)
\ \ \ \ \ \ \ \ \ 
\nvec{j} (\nvec{r},t)=\frac{1}{2 i}[\psi^{*}(\nvec{r},t)\nabla\psi(\nvec{r},t)-\psi(\nvec{r},t)\nabla\psi^{*}(\nvec{r},t)]
\label{current_eq}
\end{equation}
%

%
%
%
\section{Rydberg States in Crossed Electric and Magnetic Fields  \label{Crossed_fields_section} }
%
%
%
%
In experimental work the Rydberg atoms will often be manipulated 
by crossed perpendicular electric and magnetic fields. 
This problem has been also discussed by Pauli already in 1926 and most
of the mathematical structure was identified by Pauli already then, as is
summarized here in section \ref{pauli_el_mag_math}.

For the discussion here we  reconstruct the $n=5$  (table \ref{compare_nlm_nkm_5} ) 
with a change as table \ref{Pauli_nlm_nkm_5}. In the right hand side of 
table \ref{Pauli_nlm_nkm_5} we have interchanged the 
meaning of the quantum numbers $m$ and $k$, to adjust this to the situation
when the axis defining $k$ is x-axis and not z-axis. 
\begin{table}[htb]
\centering
\begin{tabular}{|c|c|c|c|c|c|   }      \hline
 \multicolumn{6}{|c|}{n=5}\\ \hline
 $m_z$     & \multicolumn{5}{|c|}{l  } \\ \hline
-4&    4&   &	&    &    \\  \hline
-3&    4&  3&	&    &    \\   \hline
-2&    4&  3&  2&    &    \\    \hline
-1&    4&  3&  2&   1&    \\ \hline
 0&    4&  3&  2&   1&   0\\	   \hline
 1&    4&  3&  2&   1&    \\	   \hline
 2&    4&  3&  2&    &    \\ \hline		   
 3&    4&  3&	&    &    \\ \hline
 4&    4&   &	&    &    \\ \hline
\end{tabular}
\ \ \ \ \ \ \ \ \   \   
\begin{tabular}{|c|c|c|c|c|c|c|c|c|c|  }      \hline
 \multicolumn{10}{|c|}{n=5}\\ \hline
 $k_x$     & \multicolumn{9}{|c|}{ $m_x$  } \\ \hline
-4 &\ 	&\   &\   &\   &  0 &\   &\   &\   &\	 \\  \hline
-3 &\   &\   &\   &-1  &\   & 1  &\   &\   &\	 \\   \hline
-2 &\   &\   &-2  &\   &  0 &\   & 2  &\   &\	 \\    \hline
-1 &\   &-3  &\   &-1  &\   & 1  &\   & 3  &\	 \\ \hline
 0 & -4 &\   &-2  &\   &  0 &\   & 2  &\   &  4  \\	 \hline
 1 &\   &-3  &\   &-1  &\   & 1  &\   & 3  &\	 \\	 \hline
 2 &\   &\   &-2  &\   &  0 &\   & 2  &\   &\	 \\ \hline		
 3 &\   &\   &\   &-1  &\   & 1  &\   &\   &\	  \\ \hline
 4 &\   &\   &\   &\   &  0 &\   &\   &\   &\	 \\ \hline
\end{tabular}
\caption[Stark and Spherical Basis States, $n$=5]{Stark states combinations of hydrogen atom for $n=5$ to the right, compared to
the states of spherical basis arranged in similar manner to the left\label{Pauli_nlm_nkm_5}}
\end{table}

\begin{figure}[h]
\begin{center}
\includegraphics[width= \textwidth]{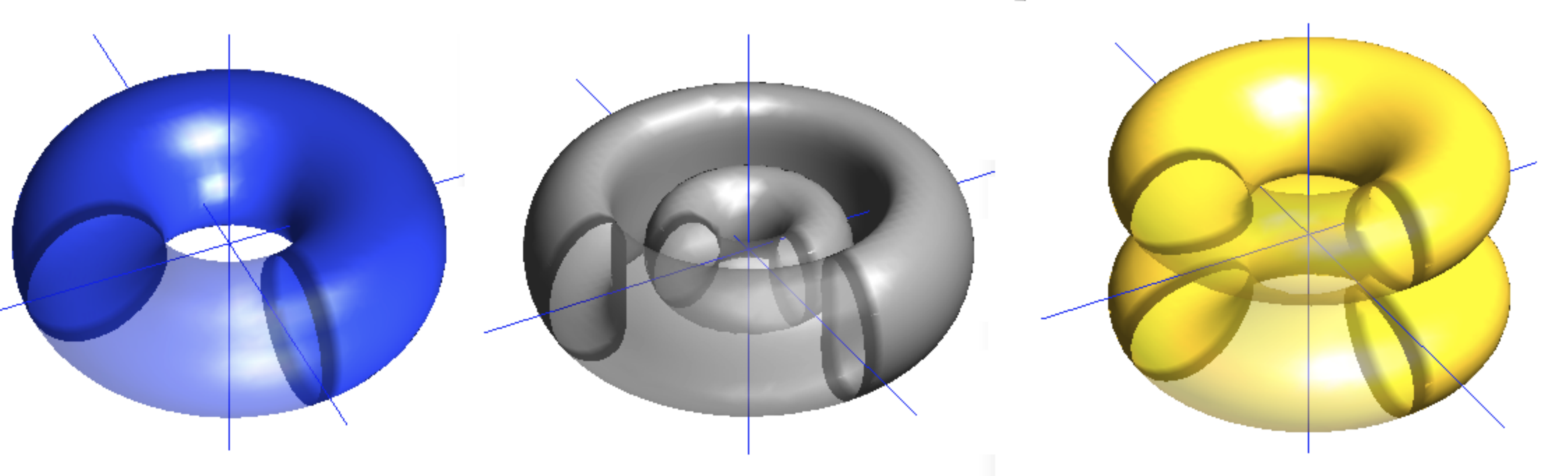}

\includegraphics[width=\textwidth]{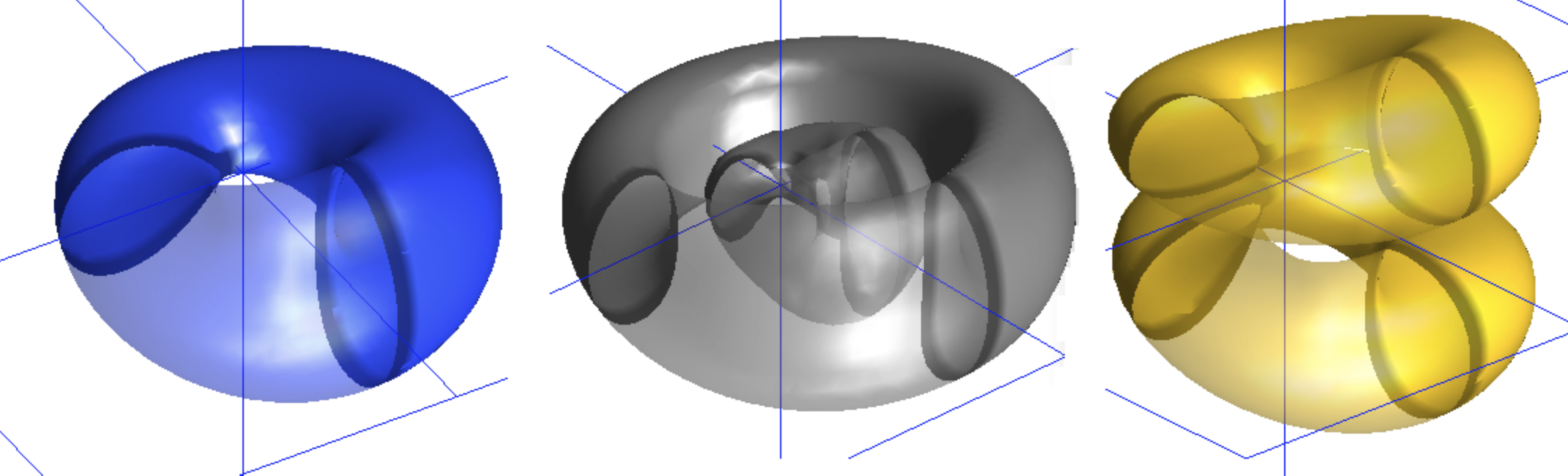} 
\includegraphics[width=\textwidth]{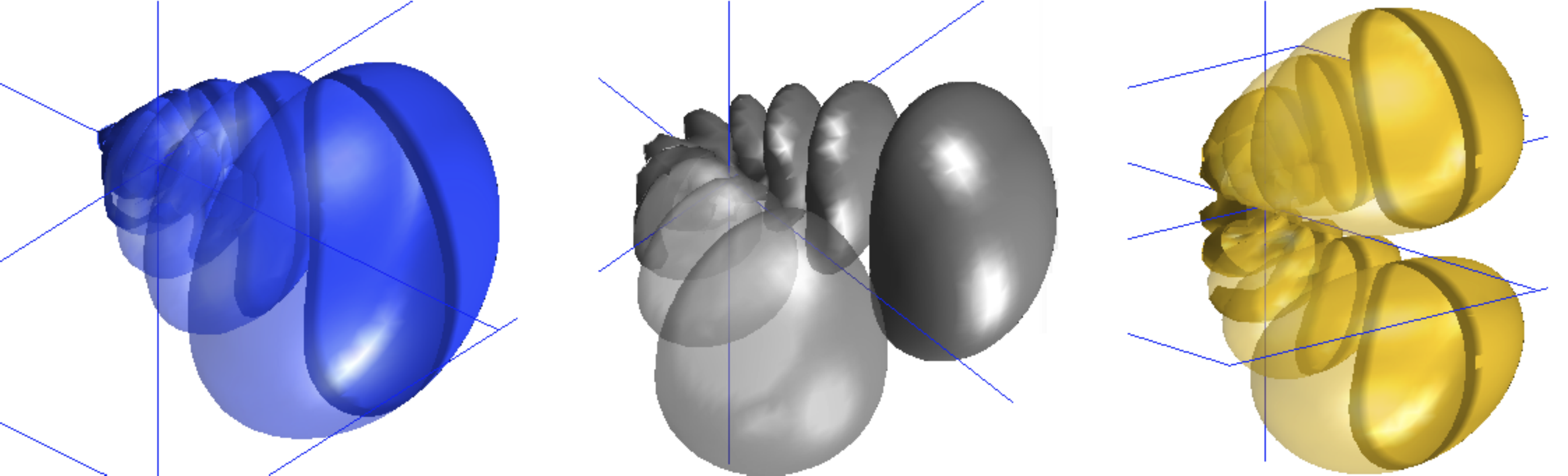}

\caption{The coherent elliptic state (in the left column) and the two quasi-coherent elliptic states 
as one changes from purely circular situation, where E is zero, via elliptic region with both 
E and B to the Stark region where magnetic field is negligible. Note that for the quasi-coherent elliptic states
the resulting Stark states are eigenstates of $A_x$ or $x$ but not  eigenstates of $L_x$. This is because the two 
states are degenerate and thus their linear combinations can be chosen. For example, the first 
two states could be combined to a $m_x=1$ and $m_x=-1$, but the connection to the
"flat states" which are determined by the magnetic field dictates the indicated shapes.
\label{CES_quasiCES2}} 
\end{center}
\end{figure}

Here we shall show some surprising features of the transition from circular states 
to linear Stark states. The Coherent elliptic state behaviour is well known. 
In figure
\ref{CES_quasiCES2} the CES is shown to the left (blue in the color print version).
The other two columns are the next two states which are degenerate 
in for all combinations of E and B as far as they remain orthogonal. 

We can see that while the CES has one ring, transforming from the circular
to the elliptic (flat structure close to the x-y plane), the two qCES have
two rings, configured in very distinct ways (inside or above each other). 
We also see that the CES develops into a cylindrically symmetric 
($m_x=0$) shape, i.e. not flat, the two qCES are transformed to two
flat objects, one close to the x-y plane, the other close to the 
perpendicular x-z plane.

\clearpage

\begin{figure}[h]
\begin{center}
\includegraphics[width= \textwidth]{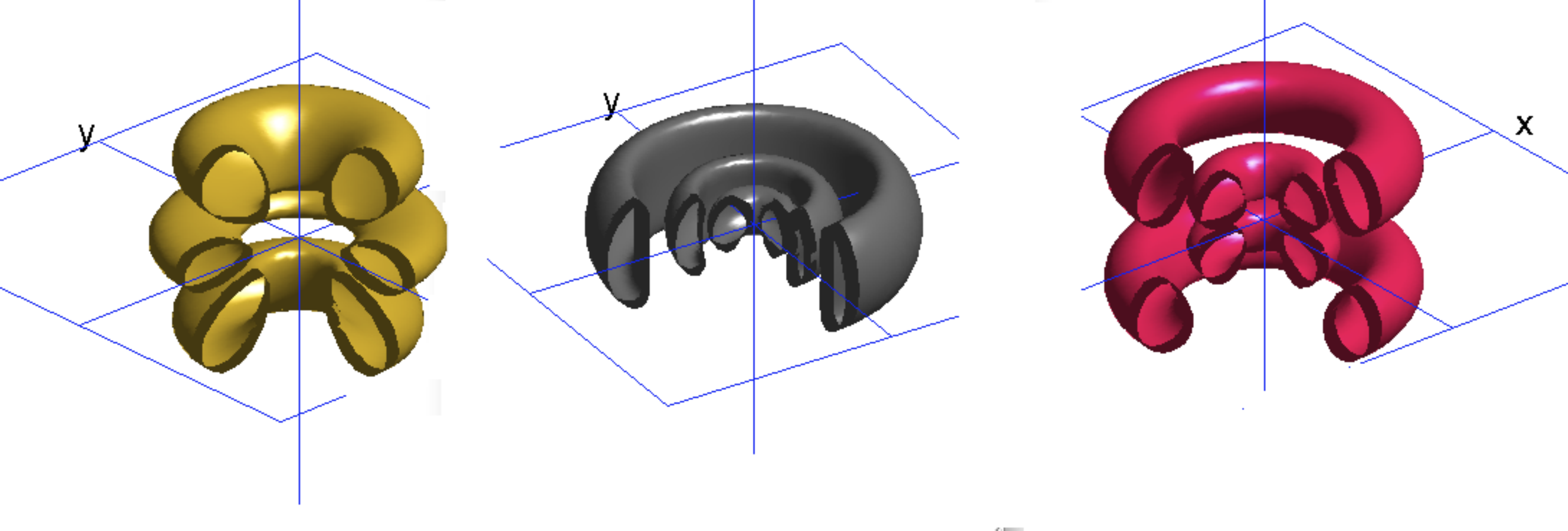}

\includegraphics[width=\textwidth]{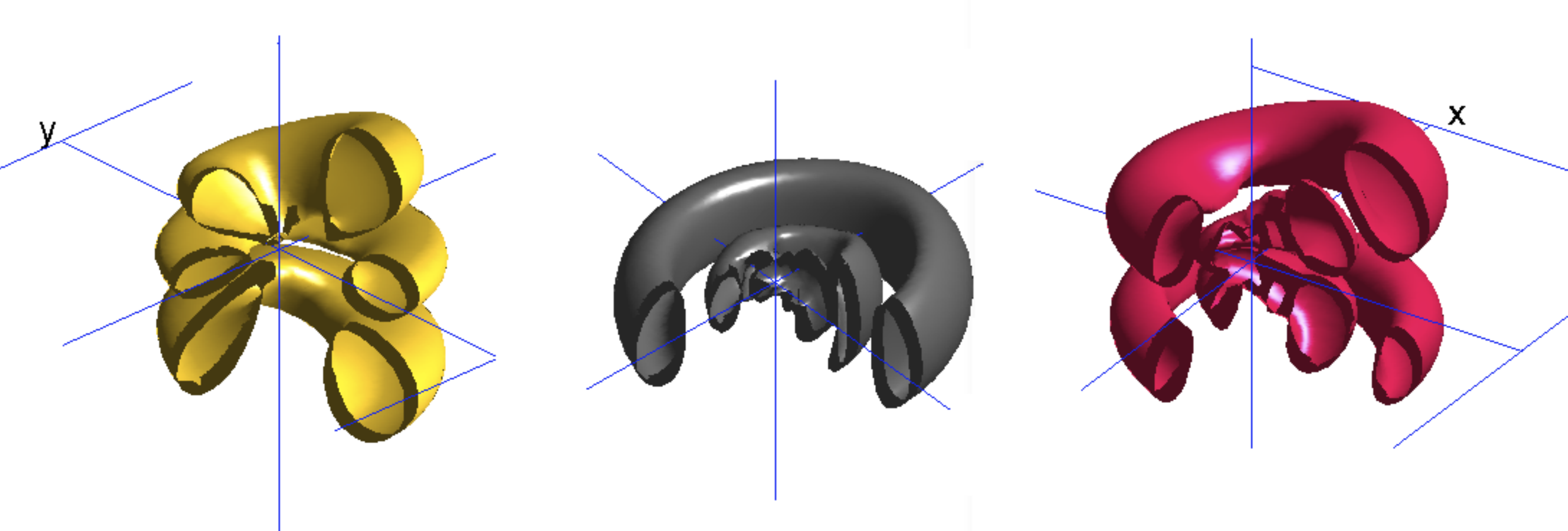} 

\includegraphics[width=\textwidth]{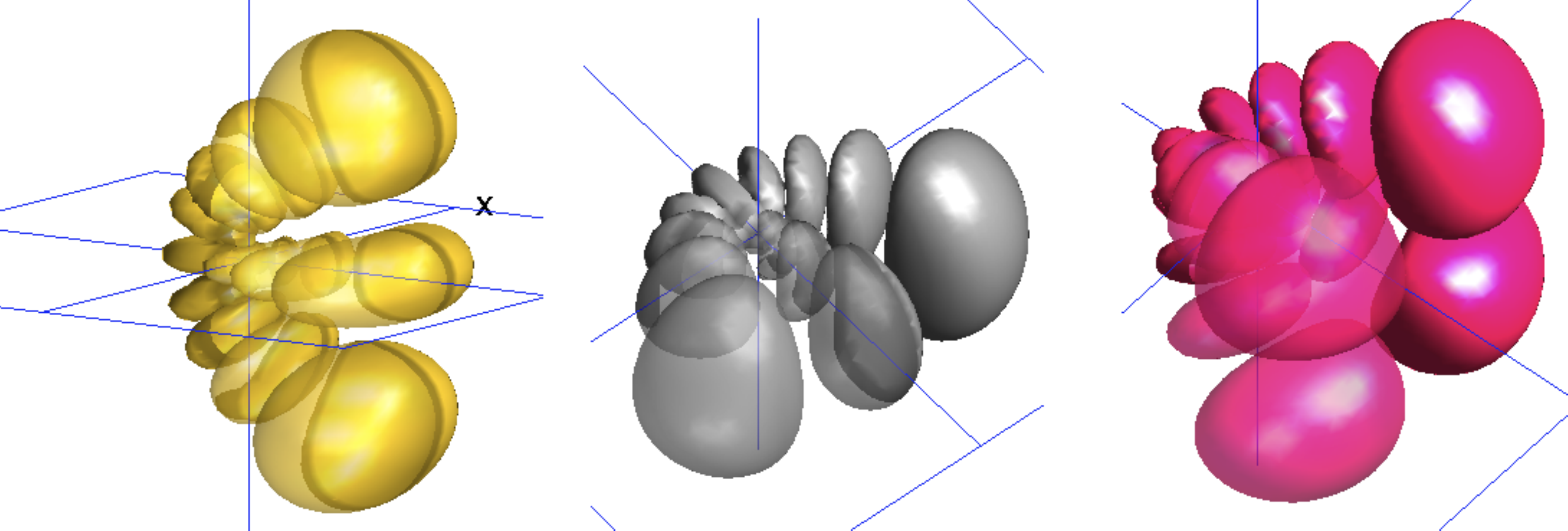}

\caption{Three quasi-coherent elliptic states of the second type which remain degenerate
as one changes from purely circular situation, where E is zero, via elliptic region with both 
E and B to the Stark region where magnetic field is negligible. Note that the resulting Stark states 
which are eigenstates of $A_x$ or $x$ are not necessarily eigenstates of $L_x$. This is because the three 
states are degenerate and thus their linear combinations can be chosen. The three 
states are $m_x=2$ and $m_x=-2$ and $m_x=0$, but the connection to the
"flat states" which are determined by the magnetic field dictates the indicated shapes.
The corresponding usually used Stark states with well defined $m$ are shown in the 
figure \ref{starks_3_deg} below.
\label{quasiCES3}} 
\end{center}
\end{figure}

\begin{figure}[h]
\begin{center}
\includegraphics[width=\textwidth]{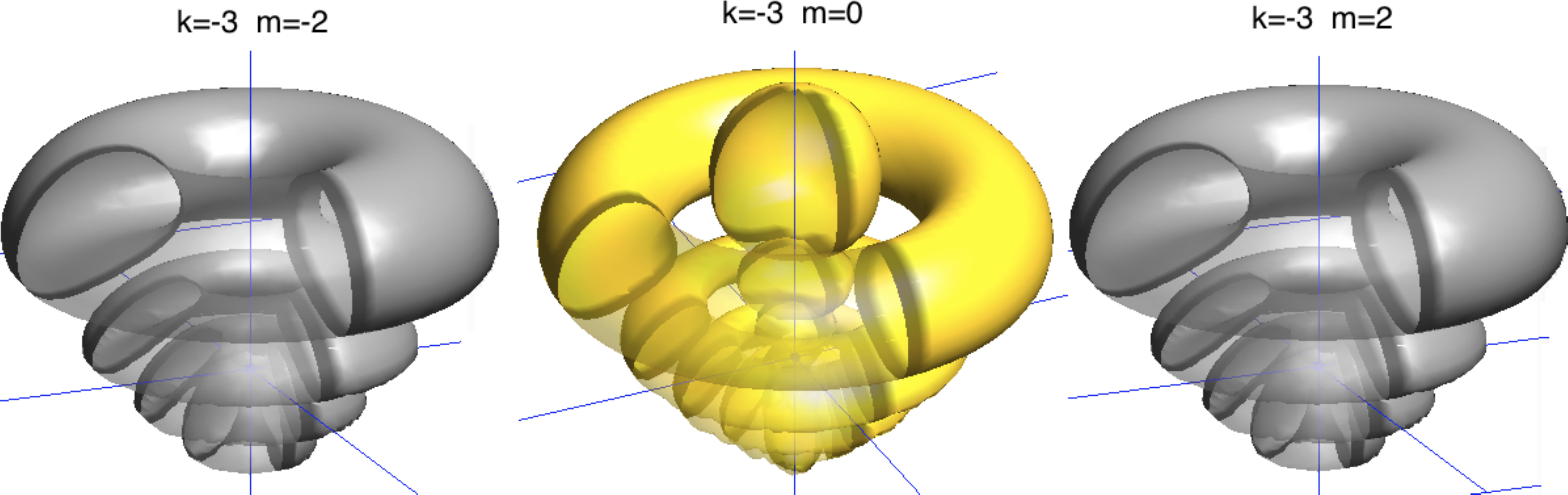}

\caption{The three stark states which combine to the three Stark states 
shown in the lowest part of the figure \ref{quasiCES3}. The middle one is $m=0$, i.e.
axially isotropic. Note that here the quantization axis is vertical as in the section on
Stark states, while the one relevant in the previous figure is horizontal.
\label{starks_3_deg}} 
\end{center}
\end{figure}

%
%
%
\section{ Conclusion   \label{Conclude_section}   }
%
%
%
%
%
%
We have shown several types of visualizations which appear useful in understanding
the shapes of various Rydberg states. In several cases found in literature the use of
modifying multiplicative factors or similar procedures introduced to hide away
unwanted features has led to a long propagation of partial misunderstandings.
We have reviewed sevaral such cases and show alternative visualizations.

In general, we have demonstrated that present day computer graphics can
give very intuitive representations of Rydberg states. 

The system used in this work is MATLAB, but very similar functionality can be found 
in many different frameworks, as e.g. Mathematica,  Python libraries, visualization toolkit VTK
and many others. The main keywords are contour plots and plots of isosurfaces for
three-dimensional distributions.

An important message of this work should also be that 
one can not expect to be able to use the standard methods with
standard parameters. Successful applications require additional research
into the abilities of the software.

\clearpage

\eject
%
%
%
%
\section*{References}
%
%
%

\end{document}